\documentclass{arxiv}
\usepackage{times}
\usepackage{epsfig}
\usepackage{graphicx}
\usepackage{amsmath}
\usepackage{amssymb}
\usepackage{color}
\usepackage{enumitem}
\usepackage{arydshln}
\usepackage{multirow}
\usepackage{bm}
\usepackage{bbding}
\usepackage{booktabs}
\usepackage{bbding}
\usepackage{amssymb} 
\usepackage{multicol}
\usepackage{adjustbox}
\usepackage{rotating} 
\usepackage[table]{xcolor}
\usepackage{array}     
\usepackage{dcolumn}    
\usepackage{siunitx}
\usepackage{tipa}

\definecolor{sblue}{RGB}{234,234,242}

\usepackage{enumitem}
\usepackage[dvipsnames]{xcolor}

\newcommand{\added}[1]{\textcolor{black}{#1}}

\usepackage[english]{babel}
\usepackage{amsthm}
\usepackage[colorlinks,linkcolor=black]{hyperref}
\newtheoremstyle{exampstyle}
{0.0em} 
{0.0em} 
{} 
{1em} 
{\bfseries} 
{.} 
{1em} 
{} 
\usepackage{balance}
\theoremstyle{exampstyle}

\usepackage{CJKutf8}


\begin{document}
	\ArticleType{REVIEW PAPER}
	\Year{2021}
	\Month{}
	\Vol{}
	\No{}
	\DOI{}
	\ArtNo{}
	\ReceiveDate{}
	\ReviseDate{}
	\AcceptDate{}
	\OnlineDate{}
	
	\title{A Literature Review of Literature Reviews \\ in Pattern Analysis and Machine Intelligence} 
	
	\author[1]{Penghai Zhao}{}
        \author[1]{Xin Zhang}{}
        \author[1]{Jiayue Cao}{}
        \author[1,2]{Ming-Ming Cheng}{}
        \author[1]{Jian Yang}{}
        \author[1,2]{Xiang Li}

	\AuthorMark{Penghai Zhao, Xin Zhang, Jiayue Cao }

    \AuthorCitation{Penghai Zhao, Xin Zhang, Jiayue Cao, et al}
	
	

	\address[1]{VCIP, CS, Nankai University, Tianjin {\rm 300000}, China}
    \address[2]{NKIARI (SHENZHEN FUTIAN), Shenzhen {\rm 518045}, China}

    \abstract{
    \added{The rapid growth of research in Pattern Analysis and Machine Intelligence (PAMI) has rendered literature reviews essential for consolidating and interpreting knowledge across its many subfields. In this work, we present a comprehensive tertiary analysis of PAMI reviews along three complementary dimensions: (i) identifying structural and statistical regularities in existing surveys; (ii) developing quantitative strategies that help researchers navigate and prioritize within the expanding review corpus; and (iii) critically assessing emerging AI-generated review systems. To support this study, we construct RiPAMI, a large-scale database containing more than 3,000 review articles, and combine narrative synthesis with statistical analysis to capture structural and content-level features. Our analyses reveal distinctive organizational patterns as well as persistent gaps in current review practices. Building on these insights, we propose practical, article-level strategies for indicator-guided navigation that move beyond simple citation counts. Finally, our evaluation of state-of-the-art AI-generated reviews indicates encouraging advances in coherence and organization, yet also highlights enduring weaknesses in reference retrieval, coverage of recent work, and the incorporation of visual elements. Together, these findings provide both a critical appraisal of existing review practices and a forward-looking perspective on how AI-generated reviews can evolve into trustworthy, customizable, and transformative complements to traditional human-authored surveys.}
    }

	
	\maketitle

\section{Introduction}
        \label{sec_introduction}

        In both natural and artificial systems, entropy—the degree of disorder—tends to increase continuously. A comparable phenomenon occurs in the realm of human knowledge. As research outputs grow rapidly, the knowledge system accumulates scattered and overlapping information, which can lead to redundancy and inefficiency in creating new insights. To counter this trend, literature reviews function as an essential organizing mechanism. Much like gravity that shaped the early universe from dispersed particles into coherent structures, reviews gather fragmented research findings and arrange them into a meaningful whole. A literature review therefore serves not only to demonstrate familiarity with the academic work on a given topic but also to position that work within a broader context. By compiling and synthesizing the most relevant publications, it provides a clear and comprehensive overview of a field and supports the efficient advancement of knowledge.
        
\begin{figure}
        \centering
        \includegraphics[width=1.0\linewidth]{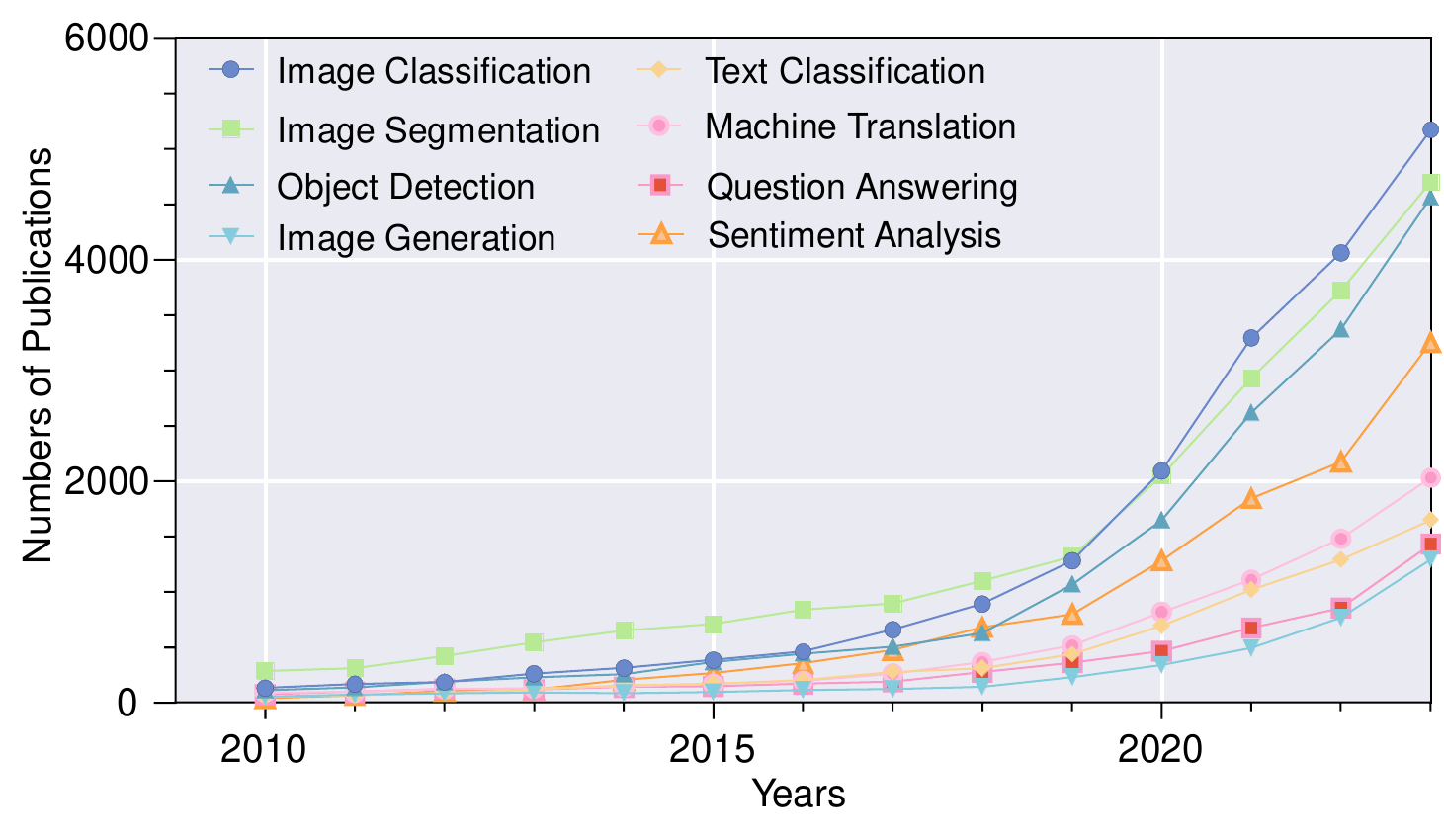}
        \caption{Annual Publication Trends of Literature Review in the Field of PAMI: A notable rising trend can be observed since 2015, reflecting an increasing scholarly focus and growing recognition of the importance of review articles in synthesizing the state of research in PAMI. The data is collected from the Google Scholar search engine.} 
        \label{fig:reviews_per_year}
    \end{figure}

        Nearly every field features its own body of literature reviews, particularly in the rapidly evolving domain of pattern analysis and machine intelligence. As a foundational technology and research direction for AI, it spans multiple areas such as image classification~\cite{chen2021review, masana2022class, machado2021adversarial,mai2022online}, image segmentation~\cite{ mazurowski2023segment, minaee2021image, siddique2021u, hao2020brief}, object detection~\cite{zaidi2022survey,cheng2023towards,qian20223d,li2020object}, natural language processing~\cite{min2021recent,hao2022recent,malik2021automatic}, etc. As reported by the AI Index Report~\cite{AIIndexReport2023}, there has been a striking surge in artificial intelligence publications, soaring from 200,000 in 2010 to nearly 500,000 by 2021. This exponential increase has subsequently led to a proliferation of related literature reviews. This trend is clearly illustrated in Fig.~\ref{fig:reviews_per_year}, which shows a marked increase in the annual publication of reviews, underscoring the growing prevalence of literature reviews in the field.

       The increasing number of literature reviews across various AI subfields has raised several concerns. First, from the authors’ perspective, the characteristics of literature reviews in the PAMI field remain underexplored, including aspects such as structural conventions, statistical patterns, and adherence to established review methodologies. Second, researchers face difficulties navigating the rapidly expanding body of literature reviews, as the sheer volume within the same field makes it harder to identify which reviews are most relevant to their research objectives. Lastly, with the growing prevalence of AI-generated reviews, it becomes crucial to assess their benefits, limitations, and overall reliability compared to traditional human-authored reviews.

        \subsection{Research Question and Contribution}

        To provide a coherent exploration of literature reviews in the PAMI field, this study is organized around three interrelated research questions that progress from current understanding, to practical strategies, and finally to future prospects:
                
        \begin{itemize}
            \item \textbf{RQ1:} \emph{\added{What structural conventions and statistical patterns characterize literature reviews in the PAMI field?}} \\
            \added{To address this question, we construct the RiPAMI database containing more than 3,000 review papers and apply both narrative synthesis and statistical analysis. Specifically, we analyze section structures, authorship patterns, reference and citation distributions, as well as visual element usage. Using LLM-based information extraction, we further quantify six representative content features (e.g., taxonomy, PRISMA, benchmark) and investigate their prevalence across sub-fields and temporal trends. This yields a comprehensive mapping of structural and statistical practices in PAMI reviews.}
        
            \item \textbf{RQ2:} \emph{\added{Given these observed patterns, how can quantitative bibliometric indicators be designed and applied to help researchers efficiently navigate and select among the growing number of reviews?}} \\
            \added{To answer this question, we design four novel, real-time, article-level, and field-normalized bibliometric indicators: the Topic Normalized Citation Success Index (TNCSI), the Impact Evolution Index (IEI), the Reference Quality Measurement (RQM), and the Review Update Index (RUI). We validate these indicators using the RiPAMI database, examining their mathematical properties, interpretability, and correlations with review impact. We further illustrate their practical utility by showing how they support indicator-guided navigation, enabling researchers to efficiently screen and prioritize reviews beyond simple citation counts.}
        
            \item \textbf{RQ3:} \emph{\added{When applied to emerging AI-generated reviews, what strengths and weaknesses can be observed compared with human-authored surveys, and what do these observations imply for their reliability and practical utility?}} \\
            \added{To answer this question, we conduct a systematic evaluation of several state-of-the-art automated review generation systems. Our analysis considers their pipeline design (e.g., intention analysis, retrieval, synthesis, report generation), reference selection strategies, structural organization, and use of visual elements. The results reveal clear patterns: recent systems are able to generate coherent and reasonably organized reviews, sometimes enriched with figures or taxonomies, yet they still suffer from critical shortcomings. These include a tendency to over-rely on highly cited but outdated references, limited capacity to recognize and integrate very recent work, and insufficient incorporation of explanatory visuals or appraisal criteria, etc. Such findings not only highlight the current gap between automated and human-authored reviews, but also underscore the importance of improving reliability, transparency, and customization if AI-generated reviews are to become practically useful in scholarly practice.}

        \end{itemize}

        In summary, we present a literature review of literature reviews and discuss the common concerns faced by existing reviews in the PAMI field. To the best of our knowledge, there has been limited scholarly attention to addressing these considerations. All the data and code framework used in this paper are publicly available at \url{https://sway.cloud.microsoft/2TXEuPuNlDKEmC9p}.

        \subsection{Organization of the Paper} 
        The remainder of the paper is organized as follows. Section~\ref{sec:methodology} briefly introduces the criteria for literature screening, methods for the RiPAMI construction, and formulas for metric calculations. Section~\ref{sec:ror} integrates narrative synthesis with statistical examination, providing both a qualitative account of structural and content patterns and a quantitative mapping of meta-data features across existing surveys. Section~\ref{sec:selecting_hq_reviews} further discusses how the proposed metrics can assist in efficiently selecting the proper reviews. The characteristics of human-authored versus AI-generated literature reviews are discussed in Sec.~\ref{sec_humanvsai}. In Sec.~\ref{sec_Challenges_Future}, we explore the challenges and future prospects of literature reviews. Finally, the paper concludes in Sec.~\ref{sec_conclusion}.

    \section{Methodology} 
    \label{sec:methodology}

        \subsection{Scope}
        This study examines all fourteen types of review articles as classified in Grant’s comprehensive typology~\cite{grant2009typology}, including narrative review, systematic review, state-of-the-art review, etc. While this research is technically a tertiary or ``umbrella'' study, we opted for the more commonly understood term ``literature review'' to improve clarity and acceptance within the PAMI field. In addition, although the quantity of reviews is considerably smaller compared to that of normal papers, it remains impractical to analyze reviews within every field. Therefore, this paper will focus only on reviews in the PAMI field.

        \subsection{Literature Selection Criteria for Narrative Reviews}
        \added{To support the analysis in Section~\ref{sec:narrative_review}, we establish specific criteria for selecting literature included in the narrative review. The search process draws on Google Scholar, Semantic Scholar, IEEE Xplore, and ScienceDirect, covering publications from major publishers such as IEEE, Elsevier, Springer, and others. }
        
        The filtering procedure follows three steps. First, we compile 106 search keywords derived from the scopes of leading journals and conferences in the PAMI domain, including terms such as \emph{speech recognition}, \emph{optical character recognition}, and \emph{self-supervised learning}. Second, we require that a paper’s title contains either “survey” or “review,” and that the chosen keyword appears in its title or abstract. Finally, we exclude preprints not published in journals or conference proceedings, ensuring that all selected articles are peer-reviewed.  
        
        These criteria provide a focused and high-quality set of review articles, which serve as the basis for the narrative analysis presented in Section~\ref{sec:narrative_review}.

    \subsection{Literature Selection Criteria for Statistical Analysis}
    \label{sec_Database}

        The immense number of literature reviews in the PAMI field emphasizes the need for approaches beyond manual selection, analysis, and synthesis. Therefore, to ensure a comprehensive and accurate analysis of literature reviews in the PAMI field, we develop an automated process that simulates the manual selection procedure to construct the RiPAMI database.

            \subsubsection{Data Source}
            \label{sec_data_src}

               \begin{table*}[ht]
	\renewcommand\arraystretch{1.1}
	\newcommand{\tabincell}[2]{\begin{tabular}{@{}#1@{}}#2\end{tabular}}
	\setlength{\fboxrule}{0pt}
	\begin{center}
    \caption{Comparisons between Various Sources: ``Counts only'' in the citations column means that the source only records the citation counts, while ``Complete'' signifies that the data source provides a complete list of citations.}
	\resizebox{0.99\textwidth}{!}{
       \begin{tabular}{c|ccccccccc}
			\hline
			Database & Title \& Authors & Venue & Abstract & Citations & References & Source Types & Free \\ 
   
			\hline\hline
   
                arXiv  & \checkmark & $\times$ & \checkmark & $\times$ & $\times$ & API-based  &  \checkmark  \\
                CrossRef  & \checkmark & \checkmark & \checkmark & Counts Only & \checkmark & API-based  &  \checkmark  \\
                Google Scholar  & \checkmark & \checkmark & \checkmark & Complete & $\times$ & Crawler-based  &  \checkmark  \\
                IEEE Xplore  & \checkmark & \checkmark & \checkmark & Counts Only & $\times$ & API-based  &  \checkmark \\
                Semantic Scholar  & \checkmark & \checkmark & \checkmark & Complete & \checkmark & API-based  &  \checkmark \\
                Web of Science  & \checkmark & \checkmark & \checkmark & Complete & \checkmark & API-based   & $\times$ \\
                Scopus  & \checkmark & \checkmark & \checkmark & Complete & \checkmark & API-based & $\times$   \\
                arXiv Data File  & \checkmark & $\times$ & \checkmark & $\times$ & $\times$ & Snapshots  &  \checkmark  \\
                CrossRef Data File & \checkmark & \checkmark & \checkmark & Counts Only & \checkmark & Snapshots  &  \checkmark  \\
                \textbf{RiPAMI\textit{(Ours)}} & \checkmark & \checkmark & \checkmark & Complete & \checkmark & Snapshots &  \checkmark \\
	
			\hline
		\end{tabular}
        \label{tab_data_src}
	}
	\end{center}
	
    \end{table*}
            Reliable data sources for analyzing extensive reviews are fundamentally important. Based on the means of data acquisition and storage, existing scientific scholar data sources may be classified into two main categories: web-based and snapshot-based sources. Web-based source data refers to the meta-data that can be retrieved from the data provider in real-time with the use of the web crawler or the API (e.g. Semantic Scholar, arXiv, CrossRef). Such an approach would only consume a small amount of storage on the local machine but needs to query meta-data every single time. On the contrary, the offline snapshot consumes a larger amount of storage space but eliminates the need for frequent API queries. Since the snapshot is a mirror image of relevant papers before a specific time, its data remains unchanged over time compared to the API-based sources. As a result, the snapshot ensures a consistent dataset in all of the experiments, avoiding issues of irreproducibility caused by changes in the provided web retrieving service.
    
            Table~\ref{tab_data_src} compares various most commonly used data sources. However, none of these sources is perfect. For example, arXiv is a free distribution service and an open-access archive for millions of scholarly articles in various fields. However, the arXiv data source fails to provide the publication venue, citations, and references. Semantic Scholar seems promising, but it suffers a lower update rate than arXiv and a narrower search scope than Google Scholar. Google Scholar is an online search engine that indexes scholarly literature from a wide range of disciplines. Although Google Scholar uses powerful automated programs to retrieve files for inclusion in its search results, it still faces challenges in many areas. Beel~\cite{beel2009google,beel2010academic,beel2010robustness} argues that Google Scholar places a high weight on citation counts in its ranking algorithm and has therefore been criticized for exacerbating the Matthew effect~\cite{merton1968matthew}. Moreover, the citation counts displayed on Google Scholar are subject to manipulation by complete nonsense articles indexed on Google Scholar (e.g. citations from AI-generated pre-print papers published on arXiv should have been ignored). Therefore, a promising engineering solution is to leverage the strengths of various approaches to overcome the weaknesses of each approach, as will be detailed in the next subsection.

        \subsubsection{Database construction} 
        \label{sec:dataset_construction}
            To ensure reproducible experiments and prevent overburdening the server, we construct an SQL-based database dubbed RiPAMI (Reviews in Pattern Analysis and Machine Intelligence, pronounced as \textipa{/ri:p\ae mi/}). This database stores information related to the paper such as title, abstract, date of publication, venue, citation counts, and reference details, etc.

            \begin{figure*}
                 \centering
                 \includegraphics[width=0.95\linewidth]{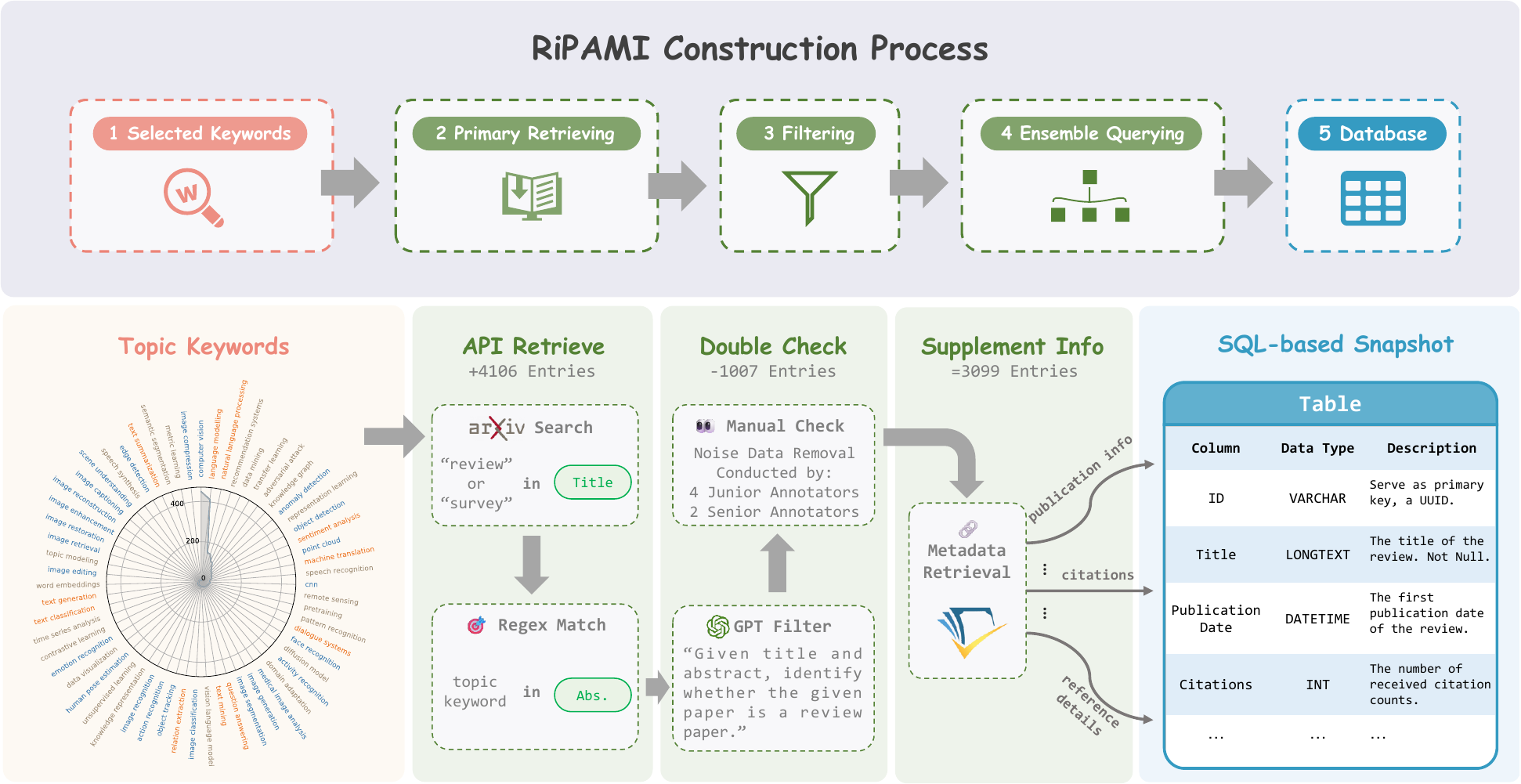}
                 \caption{\added{A Schematic Overview of the Database Construction Process: From initial keyword selection to the final SQL-based RiPAMI snapshot, three key steps are implemented to ensure the data in RiPAMI is clean, accurate, and reliable. For simplicity, the polar diagram shows only a subset of the keywords used for retrieval. Since citation counts fluctuate over time, the retrieval date for related information is set to October 2024.}}
                 \label{fig:database_construction}
            \end{figure*}

            As illustrated in Fig.~\ref{fig:database_construction}, we implement a structured process to construct the RiPAMI database, ensuring that the selected articles are review papers relevant to the pattern recognition field. The process begins with searching with keywords derived from the scopes of relevant journals and conferences \added{(check ~\ref{sec:appendix_subfields_list} for the entire keywords list)}. These keywords are then formatted into query strings for the arXiv API as follows: (ti:``review'' OR ti:``survey'') AND (ti:``{keyword.lower()}" OR abs:``{keyword.lower()}''). This structure specifically retrieves articles whose titles or abstracts include both the terms ``review'' or ``survey'' and the specified keywords. Additionally, we apply regular expressions to check that the abstracts of returned articles include the relevant keywords, ensuring that only appropriate review articles enter the RiPAMI database. To avoid potential copyright and licensing issues, papers are retrieved and downloaded using the arXiv's API. Calls to the API are made by means of HTTP requests to a certain URL. The responses will be parsed and stored in an SQL-based database. Through this process, we collect a total of 4,106 samples. To further refine our database, we employ a double-checking phase. \added{Papers are first preliminarily classified using a GPT-based filter, followed by a manual validation step in which four junior annotators conduct the initial screening and two advanced annotators provide the final confirmation to ensure the removal of false positives.} Lastly, we supplement the remaining 3,099 entries with additional metadata, including the publication date, citation counts, and reference details. As mentioned in Sec.~\ref{sec_data_src}, we suggest enriching the meta-data of papers by leveraging a combination of disparate data sources. Considering the potential legal risks of crawling to obtain academic data from Google Scholar, Semantic Scholar API was employed to obtain additional meta-data, such as citation and reference details, which are not provided by the arXiv API. The final SQL-based snapshot encompasses a wide range of data, including ID, title, publication date, citations, and more, facilitating efficient data retrieval and statistical analysis.

        \subsubsection{Meta-data Statistics of RiPAMI}
\begin{figure*}
                 \centering
                 \includegraphics[width=1.0\linewidth]{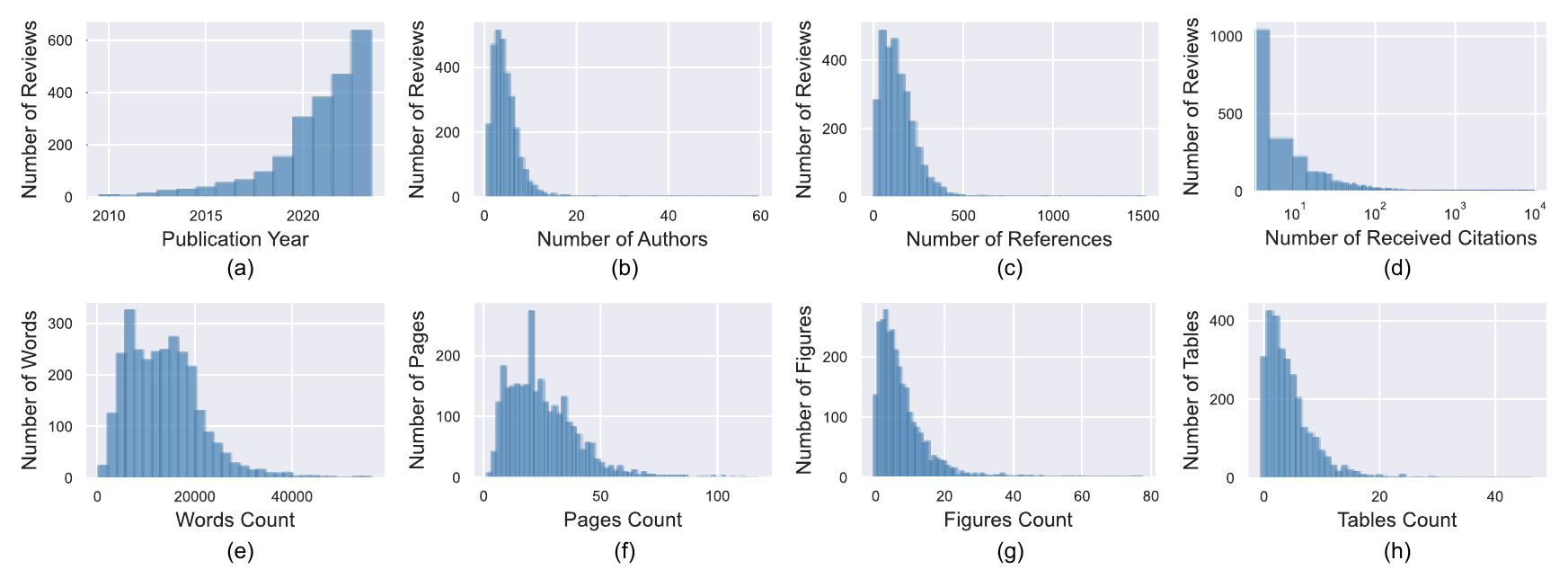}
                 \caption{\added{Statistics of the RiPAMI database: The samples in the RiPAMI database are characterized by a diversity of publication dates, scholar impact, reference numbers, etc., offering a comprehensive reflection of the state of reviews in the PAMI field.}}
                 \label{fig:stat_database}
            \end{figure*}

            The database consists of more than 3000 literature reviews from a variety of sources, publication years, and fields. To elucidate the characteristics of the RiPAMI database, we conduct a statistical analysis and plot the result in Fig.~\ref{fig:stat_database}. 
            
            \textbf{Years of Publication}
            Figure~\ref{fig:stat_database} (a) presents the distribution of publication years for literature reviews in the RiPAMI database. A clear upward trend in the number of reviews over time is evident, mirroring the pattern observed in Fig.~\ref{fig:reviews_per_year}. Both figures show a steady increase, with a notable surge between 2019 and 2020. This trend reflects a strong alignment between our sample data and the original dataset in terms of statistical consistency.
            
            \textbf{Number of the Authors}
            A review paper typically aims to provide a comprehensive overview of a specific topic. Involving multiple authors with diverse expertise could enhance the depth and breadth of the review. Fig.~\ref{fig:stat_database} \added{(b)} indicates that the majority of reviews are written by fewer than 10 authors. 
            
            \textbf{Number of the References}
            The number of references in a survey paper may influence its credibility and reliability. As shown in~\ref{fig:stat_database} \added{(c)}, the distribution of literature review references in the RiPAMI database follows a log-normal pattern. 

            \textbf{Number of the Citations}
            We count the citations of papers in RiPAMI and plot them in Fig.~\ref{fig:stat_database} (d). A power law distribution of received citations could be found.  This phenomenon where a small subset of papers receives the majority of citations is sometimes referred to as the ``Matthew Effect" or the ``Pareto principle", as reported in~\cite{merton1968matthew,brzezinski2015power}.

            \textbf{Length of Literature Reviews}
            The words count and pages number of a review article can partially reflect its depth and breadth. In general, more detailed and comprehensive reviews tend to contain a higher word count and longer pages, thus requiring more time and effort to conduct and comprehend. As shown in Fig.~\ref{fig:stat_database} (e) and (f), review articles in the PAMI field typically range from 5,000 to 20,000 words and span 10 to 40 pages. At an average reading speed of 240 words per minute~\cite{brysbaert2019many}, reading a full review would take over 20 minutes.

            \textbf{Number of the Visual Elements}
            Visual elements refer to visual representations of information, such as images and charts. They provide a clear presentation of data, simplify complex content, and enhance reader comprehension. By employing LLMs for information extraction, we may ascertain the number of figures and tables each review article contains, as depicted in Fig.~\ref{fig:stat_database} (g) and (h). It can be observed from our analysis that the majority of review articles contain fewer than 10 figures or tables.

            Further statistical details, such as maximum (Max), minimum (Min), mean, median, and mode, are available in Tab.~\ref{tab:stat_database}.
            \begin{table}[htbp]

    \centering
    \caption{Statistical Meta-data Summary of RiPAMI}
    \begin{tabular}{lccccc}
        \hline
        Attribute & Max & Min & Mean & Median & Mode \\
        \hline
        \hline
        Pub Year & 2024 & 2010 & 2021 & 2022 & 2024 \\
        Authors   & 59   & 1    & 4.61    & 4    & 3 \\
        Refs      & 1,700 & 1    & 138.19  & 120  & 66 \\
        Cites    & 11,578 & 0    & 78.48   & 11   & 0 \\
        Words    & 57,883 & 393  & 13,613.57 & 12,974 & 5,903 \\
        Pages    & 118  & 2    & 25.24   & 22   & 20 \\
        Figs     & 77   & 0    & 7.42    & 6    & 3 \\
        Tabs     & 47   & 0    & 4.38    & 3    & 1 \\
        \hline
    \end{tabular}
    \label{tab:stat_database}

\end{table}

        \subsection{Bibliometric Indicators for Reviews}

        Literature reviews reflect the state of their respective fields, highlighting current research priorities and offering a glimpse into emerging trends. Despite their significance, systematic methods for evaluating literature reviews remain underdeveloped. By introducing quantitative tools, bibliometric methods offer a promising solution, enabling researchers to objectively analyze literature reviews and uncover key patterns and trends. Bibliometrics is a research field focused on the use of quantitative methods and statistical analysis to evaluate various aspects of scholarly publications~\cite{broadus1987toward}. These methods and analysis, including widely recognized metrics like citation counts and impact factors, offer researchers valuable insights in their daily scientific work, aiding them in making prompt and informed decisions. However, most existing bibliometric methods suffer from several limitations including unfair comparison, misuse, and manipulation as introduced in the ``San Francisco Declaration on Research Assessment'' (DORA)~\cite{american2012san} and the Leiden Manifesto~\cite{hicks2015bibliometrics}. Notably, many of these metrics were not specifically designed for evaluating literature reviews, further limiting their applicability in this context.
        
         In light of these limitations, we propose two new impact indicators, $TNCSI$ and $IEI$, along with two quality indicators, $RQM$ and $RUI$.

        \subsubsection{Topic Normalized Citation Success Index (TNCSI)}
        \label{sec_tncsi}
        Most existing bibliometric methods face challenges in conducting cross-disciplinary comparisons due to variations between fields~\cite{trueger2015altmetric, purkayastha2019comparison,shen2020utilization, tong2023novel}. Although a few methods offer the capability for cross-disciplinary comparison, they either require manual assignment of field-specific keywords or involve complex and costly retrieval to calculate metrics. These drawbacks have, to some extent, hindered individual researchers from adopting these methods, thereby limiting the broader application of the metrics.
        
        To fairly compare the external influence of literature reviews across different fields within an acceptable cost, we propose guiding  Large Language Models (LLMs) to generate the topic key phrase and then calculating the success index of the current review within the same topic (see Appendix for details). The $TNCSI$ estimates the impact of research publications within an LLM-generated topic by normalizing citation counts to a scale between 0 and 1. The formula for calculating the $TNCSI$ is: 

        \begin{equation}
            TNCSI = \int_0^{\text{citeNum}} \lambda e^{-\lambda x} \,dx , x \geq 0, 
        \end{equation}
        where $\lambda$ is the scale parameter obtained through maximum likelihood estimation. Detailed steps for estimating $\lambda$ and constructing the probability density function are provided in~\ref{sec:appendix_TNCSI}.

        The $TNCSI$ demonstrates favorable mathematical properties and interpretability. First, the $TNCSI$ algorithm employs maximum likelihood estimation to convert the probability mass function into a probability density function. This process ensures that, in theory, the $TNCSI$ differentiates between papers with distinct citation counts, avoiding the assignment of identical values to them. Second, the $TNCSI$ possesses physical significance, representing the probability that a specific paper's citation count surpasses that of any other paper on the same topic. For example, a paper with a $TNCSI$ of 0.5 means it has more citations than half of the papers within the same topic. Furthermore, the calculation of $TNCSI$ is computationally efficient, as it avoids requiring a complete or precise ranking of all related papers. Instead, it provides a reliable estimation of the probability that the given paper surpasses others in terms of citation count within the same topic.
 
        \subsubsection{Impact Evolution Index (IEI)}
        \label{sec_IEI}
        Imagine a scenario where two papers, A and B, receive the same number of citations. The number of new citations per month for A remains steady, whereas the number of new citations for B grows exponentially. In this context, while acknowledging the importance of Paper A, it is generally assumed that Paper B holds a greater reference value. Analyzing the popularity or citation trends of the literature may help researchers stay informed about the latest developments and identify potential areas for future research.

        The $IEI$ is defined as the average slope across $l$ distinct points on a $n=l-1$ degree Bézier curve representing the citation trend over time. The formula for calculating \(IEI_{L_l}\) is:

        \begin{equation}
            \label{eq_IEI_avg}
            IEI_{L_l} = \sum_{a=0}^{l-1} \frac{(y_i/x_i)}{l},
        \end{equation}
        where $l$ represents the number of months observed. $x_i$ and $y_i$ are the components of the tangent vector at the $a$-th point on the Bézier curve, indicating the magnitude along the $x$- and $y$-axes, respectively. More details can be found in~\ref{sec:appendix_IEI}. 

        With the guidance of the $IEI$, researchers may further discern the various literature reviews within the same field that exhibit close citation counts. A higher $IEI$ indicates that an increasing number of studies reference the review, signaling its growing influence and attention. Therefore, in practical applications, researchers may prioritize the study of more promising reviews, as indicated by higher positive $IEI$ values, to mitigate redundancy and improve efficiency.
        
        \subsubsection{Reference Quality Measurement (RQM)}
        A literature review, in its essence, can not fabricate insights from a void. It fundamentally relies on the substance of existing references. Without a solid foundation of credible and high-quality sources, a literature review may lack the necessary building blocks to construct a meaningful analysis or argument. These sources provide the empirical evidence and theoretical context that ground the review, making the role of references indispensable in the creation of a substantial literature review.
        
        The RQM incorporates both the quality and timeliness of references by modifying the Gompertz function, which exhibits a sigmoidal growth pattern—slow at the beginning and end, with a rapid increase in the middle phase:

        \begin{equation}
            \label{eq:RQM}
            RQM = 1 - e^{-\beta \cdot e^{-(1 - ARQ) \cdot S_{mp}}}.
        \end{equation}
        
        In this equation, $\beta$ represents the shift parameter, $ARQ$ denotes the average reference quality, while $S_{mp}$ refers to the median semester count of the reference age, indicating the time from the publication date of the references to the issuance of the review (which will be detailed in ~\ref{sec:appendix_RQM}.
        
        Beyond simply assessing the average impact of references, the $RQM$ integrates references' timeliness and its influence on review quality. This metric addresses the tendency among authors to rely heavily on classic works, which may no longer fully align with current research trends. Additionally, the adjustment coefficient in $RQM$ is calibrated through statistical and optimization algorithms, tailored to the unique characteristics of each research discipline. This data-driven approach minimizes reliance on heuristic parameter settings and increases RQM's adaptability across various academic disciplines.

        \subsubsection{Review Update Index (RUI)}
        \label{sec_RUI}
        The $RUI$ refers to the measure of the extent to which a literature review is required to be updated due to the iteration of technology, theory, etc. The index is related to both the literature itself and the research interests of the topic. Generally, a high update index suggests that a literature review is in need of an immediate update. Conversely, a lower update index implies that few advances have been made to the investigated field and the review is still up-to-date. 

        The $RUI$ quantifies the necessity of updating a literature review by combining two indicators: the Coverage Difference Ratio (CDR) and the Review Aging Degree (RAD). The formula for calculating the $RUI$ is:
        \begin{equation}
            \label{eq_RUI}
            RUI = p \cdot CDR + q \cdot RAD.
        \end{equation}
        
        Here, \(p\) and \(q\) are weighting coefficients, set to 10 and 5, respectively. \added{The 2:1 ratio ensures that $CDR$ plays a dominant role, while the specific magnitudes primarily enhance numerical readability rather than altering the substantive balance between the terms.} Detailed definitions and calculations of the $CDR$ and $RAD$ are provided in the~\ref{sec:appendix_indicators}

        The proposed $RUI$ integrates both the popularity of the research field and the natural aging of individual reviews. It reflects how reviews require updating over time while taking into account the pace of progress within the field. By balancing these factors, $RUI$ offers real-time insights into how urgently a review needs updating. This urgency may arise from rapid developments in the field or the natural aging of the review’s content. $RUI$ moves beyond the simplistic reliance on publication dates, providing a more contextualized assessment of its timeliness.
        
        \subsection{Information Extraction for Reviews}
        \label{sec:IE_reviews}
        Given that this study aims to identify the common features of reviews within the field, solely relying on manual inspection would be both impractical and resource-intensive. To further investigate the features of reviews in the PAMI field, we employed LLM-based information extraction techniques to analyze the review content and parse the response to structured results. The proposed approach converts unstructured text into structured data, facilitating further analysis and interpretation. Details regarding the information extraction method can be found in the ~\ref{sec:appendix_IE_tech}.

    \section{Review of Reviews in PAMI}
        \label{sec:ror}
        \added{This section provides an integrated overview of the PAMI field by combining narrative synthesis and statistical analysis, aiming to identify common features of existing reviews and to uncover quantitative patterns across various sub-fields, thereby laying the foundation for subsequent in-depth discussion.}

        \subsection{Narrative Reviews in Various Sub-fields in PAMI}
        \label{sec:narrative_review}
        In this subsection, we offer a subjective analysis of reviews from diverse fields. Through this narrative approach, we aim to examine the composing characteristics shared by reviews across various fields. \added{It should be noted that only a subset of fields is discussed here for illustrative purposes, while the complete list of covered sub-fields is provided in~\ref{sec:appendix_subfields_list}}
        

        \subsubsection{Computer Vision}
        Computer vision is one of the most popular sub-fields of pattern recognition and machine intelligence. This section will focus on the literature reviews within the realm of computer vision.
        
        \textbf{Image Classification} refers to the task of assigning a label or a category to an input image, which is one of the most renowned tasks in the field of computer vision. A survey by Rawat~\cite{rawat2017deep} explores the development and advancements of deep convolutional neural networks (CNNs) in the field of image classification. The paper covers the historical context, their role in the deep learning renaissance, and the notable contributions and challenges faced in recent years. It highlights the remarkable progress of CNNs in image classification, while also acknowledging the ongoing research efforts to address challenges and provide recommendations for future exploration. Schmarje \emph{et al.}~\cite{schmarje2021survey} provides a comprehensive survey on semi-, self-, and unsupervised learning methods for image classification. The survey compares and analyzes 34 different methods based on their performance and commonly used ideas, highlighting the trends and research opportunities in the field. Through comprehensive analysis, the authors reveal the potential of semi-supervised methods for real-world applications and identify challenges such as class imbalance and noisy labels. Furthermore, the paper emphasizes the importance of combining different techniques from various training strategies to improve overall performance. In addition to CNNs, there exist alternative techniques for image classification. The paper~\cite{chandra2021survey} presents a comprehensive analysis of Support Vector Machines (SVM) in image classification. It discusses various techniques that can enhance classification accuracy and highlights its advancements. Liu \emph{et al.}~\cite{liu2023survey} investigate more than 100 different visual Transformers comprehensively in three fundamental CV tasks including classification, detection, and segmentation. They also propose a taxonomy to categorize various transformers into six groups. 

        \textbf{Object Detection} entails identifying and localizing objects of interest within an image or video. Liu \emph{et al.}~\cite{liu2020deep} offers a comprehensive survey on the advancements in deep learning-based generic object detection. This paper discusses an extensive range of issues, including detection frameworks, taxonomies, feature depiction, training strategies, and evaluation metrics. Though there have been significant advancements in generic object detection, the detection of small objects, which focuses on identifying objects with a small size, still presents challenges. The review conducted by Cheng \emph{et al.}~\cite{cheng2023towards} investigates 181 papers, constructs two large-scale datasets (SODA-D and SODA-A), and evaluates the performance of mainstream small object detection methods. Object detection demonstrates the utility and effectiveness across multiple domains.  Li \emph{et al.}~\cite{li2020object} and Litjens \emph{et al.}~\cite{litjens2017survey} investigate numerous methods and applications of object detection in remote sensing and medical image analysis respectively, showing that these methods have the flexibility to be applied in various scenarios and meet different needs.

        \textbf{Image Segmentation} is the process of dividing an image into meaningful and distinct regions to facilitate analysis and understanding. As described earlier, Minaee \emph{et al.}~\cite{minaee2021image} proposed a taxonomy for image segmentation methods which divides models into 11 categories. In addition to the taxonomy, the authors evaluate the quantitative performances of various methods on popular benchmarks. The paper also identifies open challenges and proposes promising research directions for future advancements in deep-learning-based image segmentation. Given that most image segmentation algorithms heavily rely on expensive pixel-level annotations, interest in weakly supervised image segmentation methods has increased. Ref~\cite{shen2023survey} surveys label-efficient deep image segmentation methods. According to the paper, weakly supervised segmentation approaches can be categorized into four hierarchical types, ranging from no supervision to inaccurate supervision. The authors investigated each of these four methods in separate sections, highlighting the strategies used to bridge the gap between weak supervision and dense prediction. Image segmentation techniques have a wide range of applications in the field of medical image processing, as introduced in~\cite{qureshi2023medical,xun2022generative,siddique2021u,ma2021loss}.
        
        \subsubsection{Natural Language Processing}

        Acclaimed as the jewel of the artificial intelligence crown, natural language processing (NLP) stands as a pivotal domain within the field of PAMI. Here, we provide a further discussion on several popular NLP research directions.
        
        \textbf{Named Entity Recognition} (NER) involves identifying and classifying named entities in text, such as person names, organizations, locations, and dates. The survey by Li \emph{et al.}~\cite{li2020survey} begins by introducing NER resources, including tagged NER corpora and off-the-shelf NER tools. Then, authors categorize existing works based on a taxonomy that considers distributed representations for input, context encoder, and tag decoder. The paper surveys representative methods for applying deep learning in various NER tasks, and provides a valuable reference for designing deep learning-based NER models. NER serves as the foundation technique for various natural language applications, such as relation extraction~\cite{nasar2021named}, knowledge graph~\cite{al2020named}, etc. Due to the linguistic variance of different languages, NER methods may also vary from language to language. Surveys about various language-specific NER could be found in~\cite{liu2022chinese,weegar2019recent,qu2023survey}.

        \textbf{Sentiment Analysis} focuses on determining the sentiment or emotion expressed in text, such as positive, negative, or neutral. Yadav \emph{et al.} introduce the process of gathering and analyzing people's opinions and sentiments from various sources such as social media platforms and blogs in their paper~\cite{yadav2020sentiment}.  The paper evaluates and compares different approaches used in sentiment analysis, with a focus on supervised machine learning methods like Naive Bayes and SVM algorithms. The common application areas of sentiment analysis and the challenges involved in accurately interpreting sentiments are also reported. The paper by Yue~\cite{yue2019survey} categorizes and compares a large number of techniques and methods from three different perspectives: task-oriented, granularity-oriented, and methodology-oriented. It also explores different types of data and advanced tools for research, highlighting their strengths and limitations. 

        \textbf{Language Modeling} involves training models to understand and generate human language. As early attempts, recurrent neural networks achieved desirable performance and wide application at that time, despite some shortcomings. The paper~\cite{yu2019review} specifically focuses on RNNs and long short-term memory (LSTM) cells. The authors highlight the limitations of traditional RNNs and emphasize the significance of LSTM in handling long-term dependencies. They discuss various LSTM cell variants and their performance on different characteristics and tasks. Furthermore, the paper also categorizes LSTM networks into two major types: LSTM-dominated networks which optimize connections between inner LSTM cells, and integrated LSTM networks which incorporate advantageous features from various components. Recently, LLMs have drawn widespread attention.  LLMs demonstrate significant performance improvements and unique abilities such as in-context learning, setting them apart from smaller-scale models. By investigating more than 600 works, Zhao \emph{et al.}~\cite{zhao2023survey} conduct a comprehensive review of the recent advancements in LLMs. The authors discuss the evolution of language modeling techniques, from statistical models to neural models, and highlight the emergence of pre-trained language models as a powerful approach in NLP tasks. The survey focuses on LLMs with a parameter scale exceeding 10 billion and explores four key aspects: pre-training, adaptation tuning, utilization, and capacity evaluation. The paper also presents available resources for developing LLMs and discusses important implementation guidelines. Overall, this survey serves as an up-to-date and valuable reference for researchers and engineers interested in the field of LLMs.
        
        \subsubsection{Others}
        Reviews in other popular sub-fields will be investigated in this section. 
        
        The paper by Zhou \emph{et al.}~\cite{zhou2023comprehensive} covers the evolution of pre-trained foundation models from BERT to ChatGPT and highlights their significance as parameter initializations for downstream tasks. The survey explores popular pre-trained foundation models in text, image, and graph modalities, discussing their components, pre-training methods, and advancements thoroughly. The paper also addresses topics including model efficiency, compression, security, and privacy, while offering valuable insights into scalability, logical reasoning ability, and cross-domain learning. Another survey paper~\cite{yu2023self} presents a comprehensive review of the state-of-the-art in self-supervised recommendation (SSR). The paper proposes an exclusive definition of SSR and develops a taxonomy that categorizes existing SSR methods into four categories: contrastive, generative, predictive, and hybrid. It further introduces an open-source library called SELFRec, which incorporates a wide range of SSR models and benchmark datasets. Through rigorous experiments and empirical comparison, the paper derives significant findings related to the selection of self-supervised signals for enhancing recommendation. The conclusion highlights the limitations and outlines future research directions in the field of self-supervised recommendation.

        \subsection{Popular Structure of Reviews in PAMI}
        \label{sec:review_structure}
\begin{figure}
             \centering
             \includegraphics[width=0.7\linewidth]{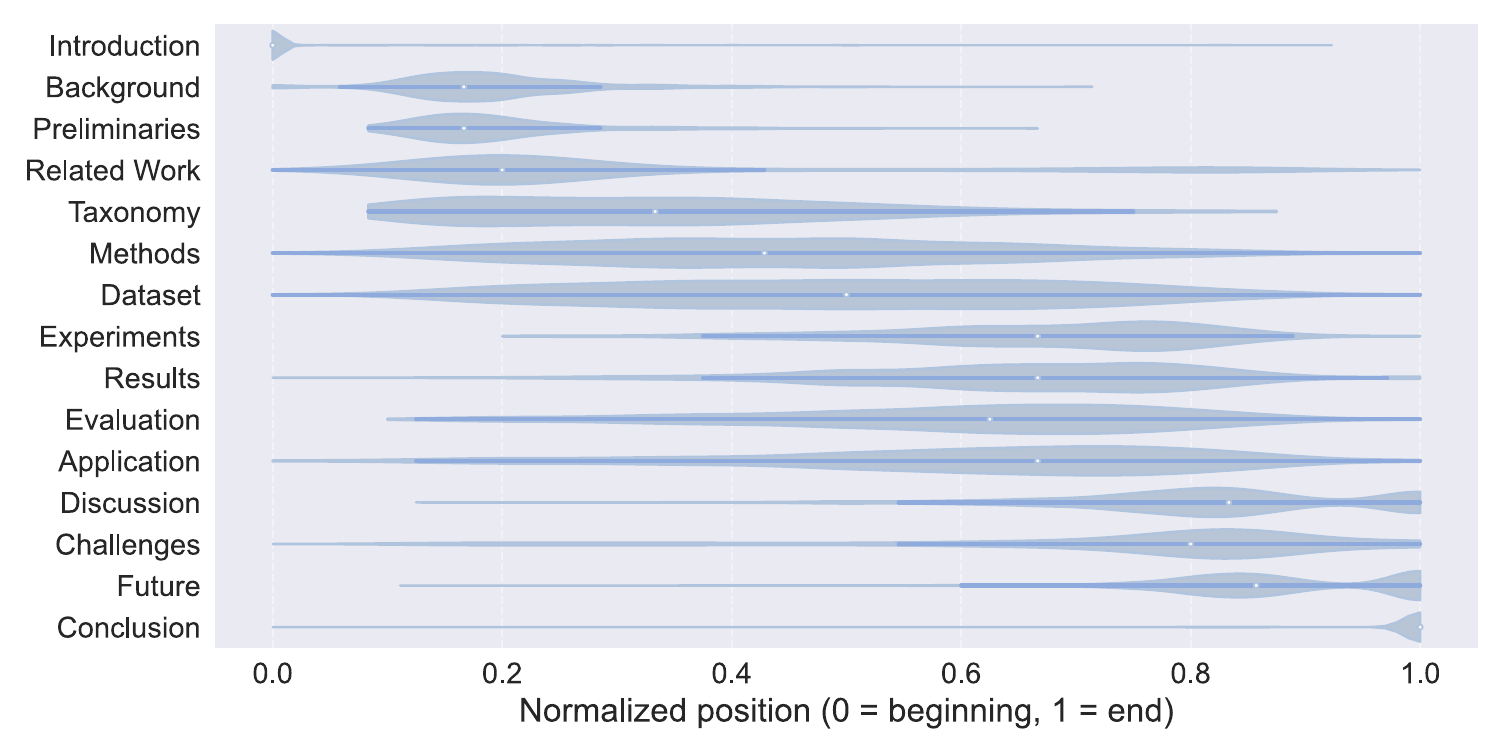}
             \caption{\added{Distributions of normalized section positions across survey papers: Core sections such as Introduction and Conclusion exhibit highly consistent placement at the beginning and end, while methodological, empirical, and reflective components show broader variation across the document body.}}
             \label{fig:review_structure}
         \end{figure}

        \added{The review's structure is alternatively referred to as the framework of the review. It is the outline that authors consider when starting to conduct the survey. Generally, a well-designed structure is considered essential for enhancing a paper’s readability and helping readers grasp its core concepts and knowledge. Such a framework typically includes foundational components such as the abstract, introduction, methodology, discussion, and conclusion, each serving a distinct communicative function. To investigate how these elements are actually organized, we conducted a large-scale analysis of section titles across our collected corpus of survey papers. We first identified a representative set of section keywords (e.g., Introduction, Methods, Experiments, Results, Discussion, Future Work, Conclusion) and extracted their occurrences. For each keyword, we normalized its position within the document (ranging from 0 at the beginning to 1 at the end) and then aggregated these values across all papers. The resulting distributions, illustrated with violin plots in Fig.~\ref{fig:review_structure}, reveal clear structural anchors: ``Introduction'' consistently appears at the very beginning, while ``Conclusion'' is concentrated at the end. Expository sections such as ``Background'' and ``Preliminaries'' occur early; ``Related Work'' and ``Taxonomy'' appear in the early-to-middle part; and ``Methods'' and ``Dataset'' are most often located around the middle. Empirical sections (``Experiments'', ``Results'', ``Evaluation'') tend to cluster in the latter half, whereas reflective components (``Application'', ``Discussion'', ``Challenges'', ``Future'') are predominantly placed near the end. More detailed information on the algorithm can be found in~\ref{sec:appendix_pos_stat_sec_titles}.}

        To gain a deeper understanding of structural practices, we applied information extraction techniques to the RiPAMI database. Using an ensemble query approach based on LLMs, we automatically identified six representative content features: (1) Taxonomy — whether the review explicitly outlines a taxonomy of methods; (2) PRISMA — whether it specifies inclusion/exclusion criteria akin to PRISMA guidelines~\cite{page2021prisma}; (3) Preliminary — whether it provides background knowledge in a distinct section; (4) Benchmark — whether it presents quantitative comparisons across methods; (5) Application — whether it discusses real-world applications; and (6) Discussion — whether it addresses challenges, limitations, or future directions. Each feature was recorded in binary format (0/1) for statistical analysis (see~\ref{sec:appendix_IE_tech} for extraction details).
        
\begin{figure}
             \centering
             \includegraphics[width=1.0\linewidth]{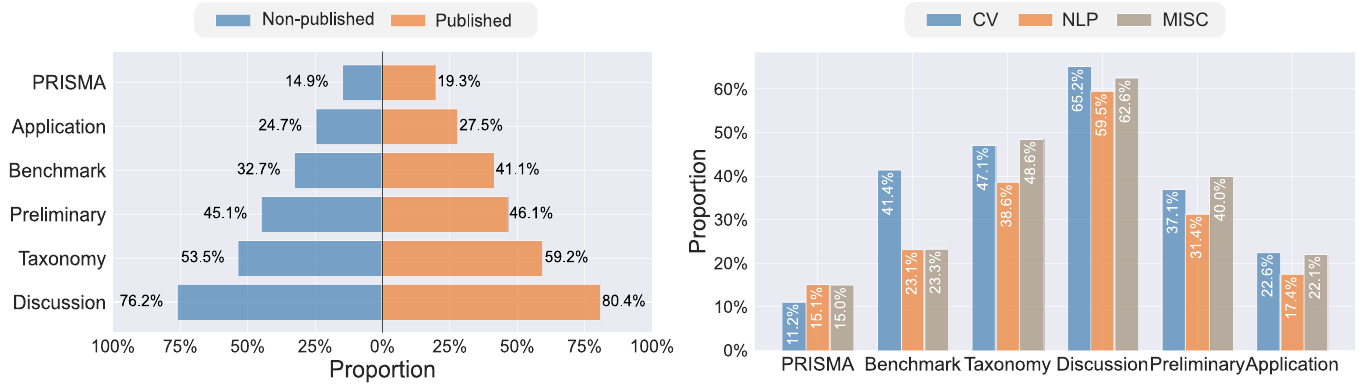}
             \caption{Statistical Analysis of Review Features in the RiPAMI Database using LLM-based Information Extraction. “Published” refers to peer-reviewed review papers, whereas “Non-published” denotes preprints without peer review. CV, NLP, and MISC correspond to the fields of computer vision, natural language processing, and miscellaneous domains, respectively.}
             \label{fig:review_feature_stat}
         \end{figure}

        The left panel of Figure~\ref{fig:review_feature_stat} presents the overall prevalence of the six identified features. A majority of reviews include a dedicated discussion section on challenges and future directions, reflecting the academic expectation that surveys should not only summarize existing work but also identify open problems and potential avenues for research. In contrast, fewer than 20\% of reviews in the PAMI field adopt PRISMA-like methodological criteria, regardless of publication venue, revealing substantial room for more rigorous systematic practices. \added{The right panel of Figure~\ref{fig:review_feature_stat} highlights domain-specific differences: reviews in computer vision (CV) show a pronounced emphasis on benchmarking, whereas those in smaller or cross-disciplinary areas (MISC) more frequently incorporate preliminaries.}

\begin{figure}
             \centering
             \includegraphics[width=0.5\linewidth]{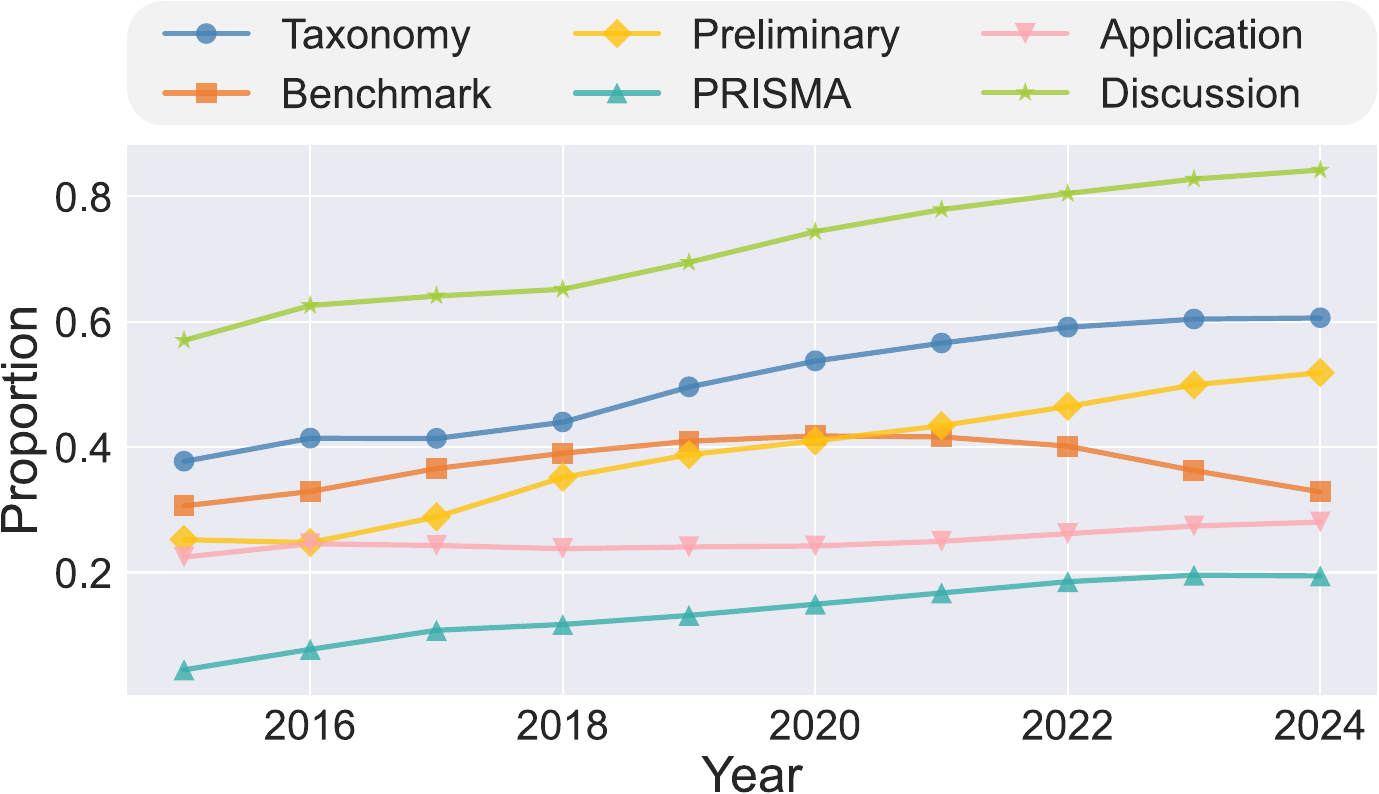}
             \caption{Trends in the Proportions of Review Features over Time: The proportion of most features has been rising progressively over the years. For improved visual clarity, Gaussian smoothing is applied to the raw data.}
             \label{fig:review_feature_change_stat}
         \end{figure}

        Figure~\ref{fig:review_feature_change_stat} illustrates temporal trends. All six features have increased steadily over the past decade, with the inclusion of Discussion'' and Preliminary'' sections showing the sharpest rise. Particularly noteworthy is that the adoption of PRISMA-style reporting has grown by more than 10\% between 2015 and 2024, pointing to a gradual shift toward more structured and methodologically robust reviews within the PAMI community.

        \added{Building on the preceding analysis, we now discuss the four major components commonly observed in survey papers:  the abstract, the beginning part, the middle part, and the ending part.}
         
        \subsubsection{Abstract}
        The abstract is widely regarded as the most important part of a paper, as it provides a concise summary of the core content, enabling readers to quickly grasp the main points and conclusions. While the level of detail may vary, a structured abstract improves clarity and reproducibility by dividing the content into clearly labeled sections, each fulfilling a specific communicative function. Importantly, international reporting standards such as the PRISMA guidelines~\cite{page2021prisma} explicitly recommend the use of structured abstracts in systematic reviews and meta-analyses, as they enhance transparency and facilitate information retrieval~\cite{kostoff2002open,beller2013prisma}. A typical structured abstract consists of sections such as Background, Objective, Methods, Results, Conclusion/Discussion, and sometimes Limitations. For instance, Sariyanidi \textit{et al.}~\cite{sariyanidi2014automatic} structured their abstract by introducing the background on facial affect, highlighting persistent challenges in the field, and then summarizing their methods, results, and discussion.

\begin{figure}
             \centering
             \includegraphics[width=1.0\linewidth]{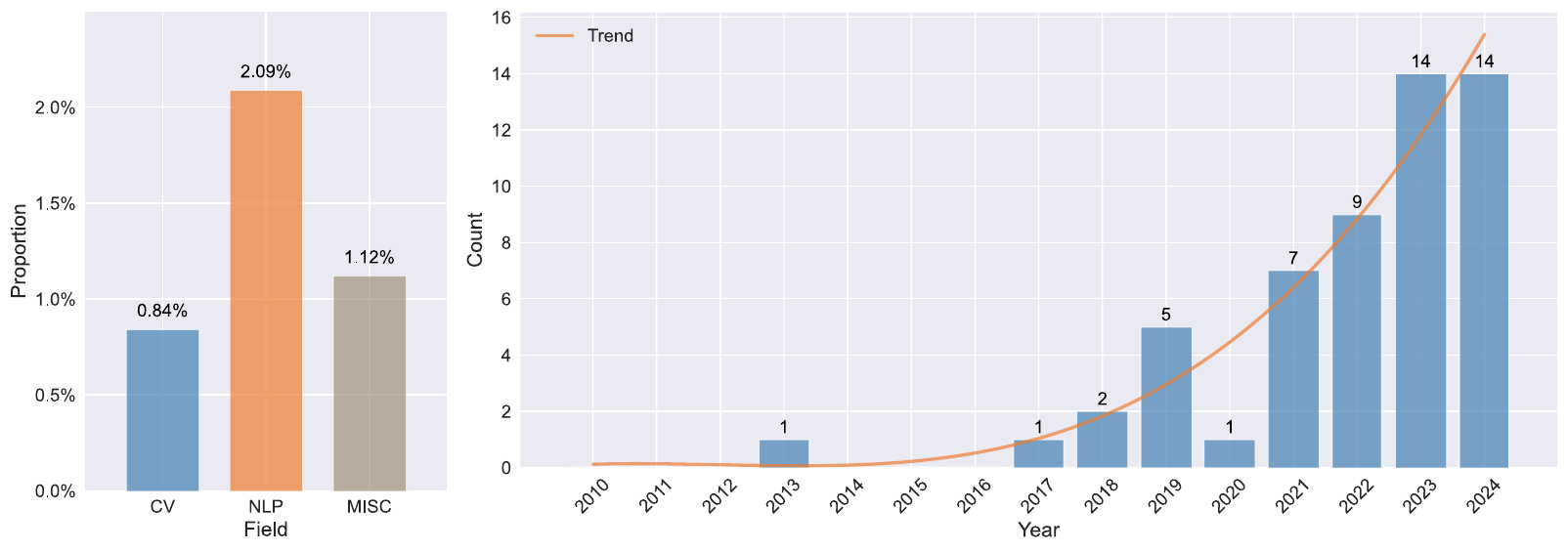}
             \caption{\added{Proportion and Trend of Structured Abstracts: The bar chart on the left shows the proportion of structured abstracts across different fields (CV, NLP, MISC), while the line chart on the right illustrates the trend in the number of structured abstracts over the years from 2010 to 2024.}}
             \label{fig:structure_abs_ratio}
         \end{figure}

        \added{We further instruct the LLM to assess whether an abstract adheres to the structured format outlined by PRISMA guidelines (detailed in ~\ref{sec:appendix_IE_tech}). As shown in Fig.~\ref{fig:structure_abs_ratio} (a), the proportion of structured abstracts in the PAMI field remains relatively low, suggesting that this field has yet to fully embrace the structured abstract format. However, panel (b) illustrates a clear upward trend in the number of abstracts meeting the structured abstract requirements, with a steady increase observed year on year. This highlights a growing adoption of standardized abstract formats in recent research.}

        \subsubsection{The Beginning Part}
        Similar to the research article, the introduction section of the literature review usually includes contextualizing the research topic, identifying the knowledge gap, defining the scope and objectives, and laying the foundation knowledge for wider audiences. ``Introduction'' usually lies at the very beginning of the paper. Most reviews first introduce the definition of the topic to acquaint readers with a basic understanding of the field. For instance, the term ``Named Entity Recognition'' may be unfamiliar to many scholars outside the field of natural language processing. A clever introduction might provide the origins of the terminology and inform the readers what is named entity recognition, as has been done in Ref~\cite{Nadeau2007ASO}. A brief definition of the field in the introduction is also welcomed for some relatively popular fields, e.g., the literature review~\cite{rawat2017deep} examines the concept of image categorization in the first sentence.

        In addition to contextualizing the research topic, many introductions also highlight the existing research and identify gaps or challenges in the current understanding of the topic. It sets the stage for the literature review by explaining why the research being reviewed is necessary and what gaps are expected to be filled. Wang \emph{et al.}~\cite{wang2022self} emphasize that while there are comprehensive surveys of self-supervised learning in computer vision, there is a lack of a similar overview specifically tailored to the remote sensing community. 
        
        Readers need to evaluate if the article is worth reading before investing more time in it. A statement of the scope or contributions of the article is a way for readers to quickly assess its relevance. In the first section, Wangkhade \emph{et al.}~\cite{wankhade2022survey} provide us with several important contributions including analyzing well-known technologies, proposing taxonomies of approaches, and summarizing benefits and challenges of sentiment analysis.

        Preliminaries and problem formulations are also popular with readers, as they both provide profound background knowledge. Given that the Gumbel-max involves considerable mathematical concepts and calculations, a ``Preliminaries'' section in the review of the Gumbel-max trick~\cite{huijben2022review} serves the purpose of providing background information and basic understanding related to the Gumbel-max trick. 

        For systematic reviews, a clear statement regarding the review methodology (e.g., literature search strategy and criteria for inclusion and exclusion) is essential~\cite{page2021prisma,kitchenham2010systematic}. Such a statement helps to minimize potential bias and enables other researchers to replicate the study's findings. Memon \textit{et. at.}~\cite{memon2020handwritten} dedicate a section to detailing their review process, including subsections on the review protocol, inclusion and exclusion criteria, search strategy, study selection process, quality assessment criteria, and data extraction and synthesis, ensuring transparency and rigor in their methodology.

        \subsubsection{The Middle Part}
        
        The middle part of a survey paper, also known as the review part, presents a detailed examination of relevant research studies, methodologies, findings, and theories related to the chosen theme. Beyond mere description, it synthesizes information from the literature to provide a cohesive and integrated understanding of the topic. This synthesis not only aligns disparate works but also evaluates their contributions and interrelations. \added{It often includes comparative analysis, highlighting similarities and differences in approaches, and may identify limitations in the existing research that warrant further investigation.}
        
        \added{Across existing surveys in the PAMI field, different organizational strategies can be observed. Some reviews arrange the discussion by grouping methods according to their technical characteristics, thereby creating typologies that help readers compare families of approaches. For instance, Minaee \emph{et al.} conducted a comprehensive survey on deep learning-based image segmentation models~\cite{minaee2021image}, grouping them into ten categories such as fully convolutional models, encoder–decoder-based models, and attention-based models. This mode of presentation highlights the breadth of methodological options, but can be less accessible for readers who are not yet familiar with the field.}
        
        \added{Other surveys are organized around specific research problems or tasks, with each section devoted to a challenge and the solutions proposed for it. A review on egocentric vision hand analysis by Bandini \emph{et al.}~\cite{bandini2020analysis} illustrates this approach: the paper is divided into sections such as hand segmentation, hand detection, and hand identification, with subsections that further explore issues like robustness to illumination changes or lack of pixel-level annotations. This problem-driven framing can be especially useful in application-oriented areas, though it may offer less systematic elaboration of theoretical underpinnings.}  
        
        \added{In many cases, authors blend these strategies. A review may be structured by tasks or challenges, while within each task methods are further grouped and compared. Guo \emph{et al.}~\cite{guo2020deep}, for example, divided their survey into task-based sections but introduced similar approaches group by group within each task. Such a blended organization can provide both problem-driven context and methodological coverage, giving readers multiple perspectives on the same body of work.}  
        
        \added{Overall, the middle part of a survey is where synthesis and comparison take place, and where authors draw together diverse contributions into a structured narrative. Whether arranged around methods, challenges, or a combination of both, these organizational choices illustrate how surveys strive to balance comprehensiveness, accessibility, and clarity in guiding readers through complex literature.}

        \subsubsection{The Ending Part}
        The subsequent ending part serves as a succinct conclusion to the reviewed studies. Within this section, the authors prefer to sum up the key findings to answer the question in the beginning part. It usually highlights both the advantages and disadvantages of the reviewed literature to conclude research gaps and suggest potential avenues for future research. 

        \added{One of the main objectives of the concluding part is to provide a concise summary of the key insights derived from the reviewed studies and to emphasize their significance and relevance to the research topic. In some literature reviews, however, the analysis and synthesis are so extensive that the main body spans a considerable number of pages. For instance, Liu et al.~\cite{liu2020deep} present a survey whose main body covers 51 double-column pages, posing a challenge for readers who may not have the time to read it thoroughly. Fortunately, the paper also provides a concise summary, which enables readers to quickly grasp the essential content and navigate to sections of particular interest.}

        When discussing further in the ending part, the authors may attempt to identify gaps and point out future directions through the analysis of existing literature. Some of the papers present the gaps and future directions separately. Hassain \emph{et al.}~\cite{hussain2021comprehensive} first discussed major challenges of multi-view video summarization such as lack of synchronization, instability of camera, and crowded scenes individually. Where there are challenges, there are future research directions. In addition to challenges, the authors also provide recommendations and future directions from various perspectives including models, benchmark datasets, and agents-based MVS, etc. Conversely, some papers choose to combine discussions of current issues with emerging research trends, as seen in the 'Future Directions' section of Ref\cite{yang2022survey}.
    
        For literature reviews delving into pragmatic methodologies, the inclusion of an applications section is quite fitting. Ref~\cite{li2021survey} offers a comprehensive review of various surveys on convolutional neural networks, dedicating a distinct section to the typical applications of 1-D, 2-D, and multidimensional CNNs. Similarly, Ref~\cite{LU2023109480} illustrates the deployment of few-shot learning techniques across disciplines like computer vision, natural language processing, and reinforcement learning.

        \section{Indicator-Guided Navigation for Literature Reviews}
        \label{sec:selecting_hq_reviews}

        \added{As the number of published reviews continues to grow, it has become common for multiple surveys to cover similar areas, attempt to answer analogous research questions, and often arrive at comparable conclusions. This overlap creates information redundancy and, to some extent, reduces research efficiency. Human scholars frequently rely on academic search engines such as Google Scholar, Semantic Scholar, and Web of Science, prioritizing high-ranking results typically sorted by citation counts to identify a select number of reviews for closer examination. However, these rankings are sometimes perceived as biased or even manipulated~\cite{ibrahim2024google,martin2021google}, and an over-reliance on citation counts can intensify the Matthew effect, which may ultimately hinder sustainable academic development within the field.}
        
        \added{Building on the indicators introduced in Sec.~\ref{sec:methodology}, this section demonstrates how they can be applied to enable indicator-guided navigation. Such navigation not only helps researchers more effectively evaluate and select reviews, but also provides a principled mechanism for AI-for-Research systems that use human-authored surveys as starting points for automated outline generation and knowledge synthesis. In this way, the indicators serve as a bridge between methodological rigor and practical strategies for navigating the expanding corpus of reviews.}

        \subsection{Appraising Reviews across Various Topics}
        A fundamental need may lie in the assessment of reviews across various fields, as most reviews are primarily aimed at readers who are relatively new to a field and may lack an in-depth understanding of the field. For instance, a researcher working in a popular field like object detection may start exploring a less familiar subfield such as optical character recognition (a subfield of computer vision). When encountering a review with only 100 citations, the researcher might dismiss its importance, unaware that in a niche area, 100 citations can actually indicate significant influence. This could lead to an oversight of critical insights in the subfield, simply due to differing citation norms between areas. The proposed $TNCSI$ offers an effective solution by automatically identifying the field of a given article and dynamically estimating its normalized impact based on the field and its current citation count. Illustrative examples are presented in Tab.~\ref{tab:ex_TNCSI}.
        
        \begin{table}[ht]
    \renewcommand\arraystretch{1.1}
    \setlength{\tabcolsep}{6pt}
    
    \begin{center}
        \caption{Example Use Cases of $TNCSI$: A similar number of citations (Cites) in various topics may lead to notably different $TNCSI$ values.}
        \begin{tabular}{m{0.4\linewidth}m{0.1\linewidth}m{0.18\linewidth}m{0.12\linewidth}}
            \hline
            Title  & Cites & Topic & $TNCSI$↑ \\
            \hline
            \hline
            Involvement of Machine Learning for Breast Cancer Image Classification: A Survey~\cite{nahid2017involvement} & 104 & Breast Cancer Image Classification & 0.79 \\
            \hline
            Class-Incremental Learning: A Survey~\cite{zhou2024class} & 99 & Class-Incremental Learning & 0.61 \\
            \hline
            A Systematic Survey of Prompt Engineering on Vision-Language Foundation Models~\cite{gu2023systematic} & 100 & Prompt Engineering & 0.94 \\
            \hline
            End-to-End Speech Recognition: A Survey~\cite{prabhavalkar2023end} & 100 & End-to-End Speech Recognition & 0.59 \\
            \hline
        \end{tabular}
        \label{tab:ex_TNCSI}
    \end{center}
\end{table}

        \subsection{The Role of Reference Quality in Selecting Reviews}
\begin{figure}
             \centering
             \includegraphics[width=0.7\linewidth]{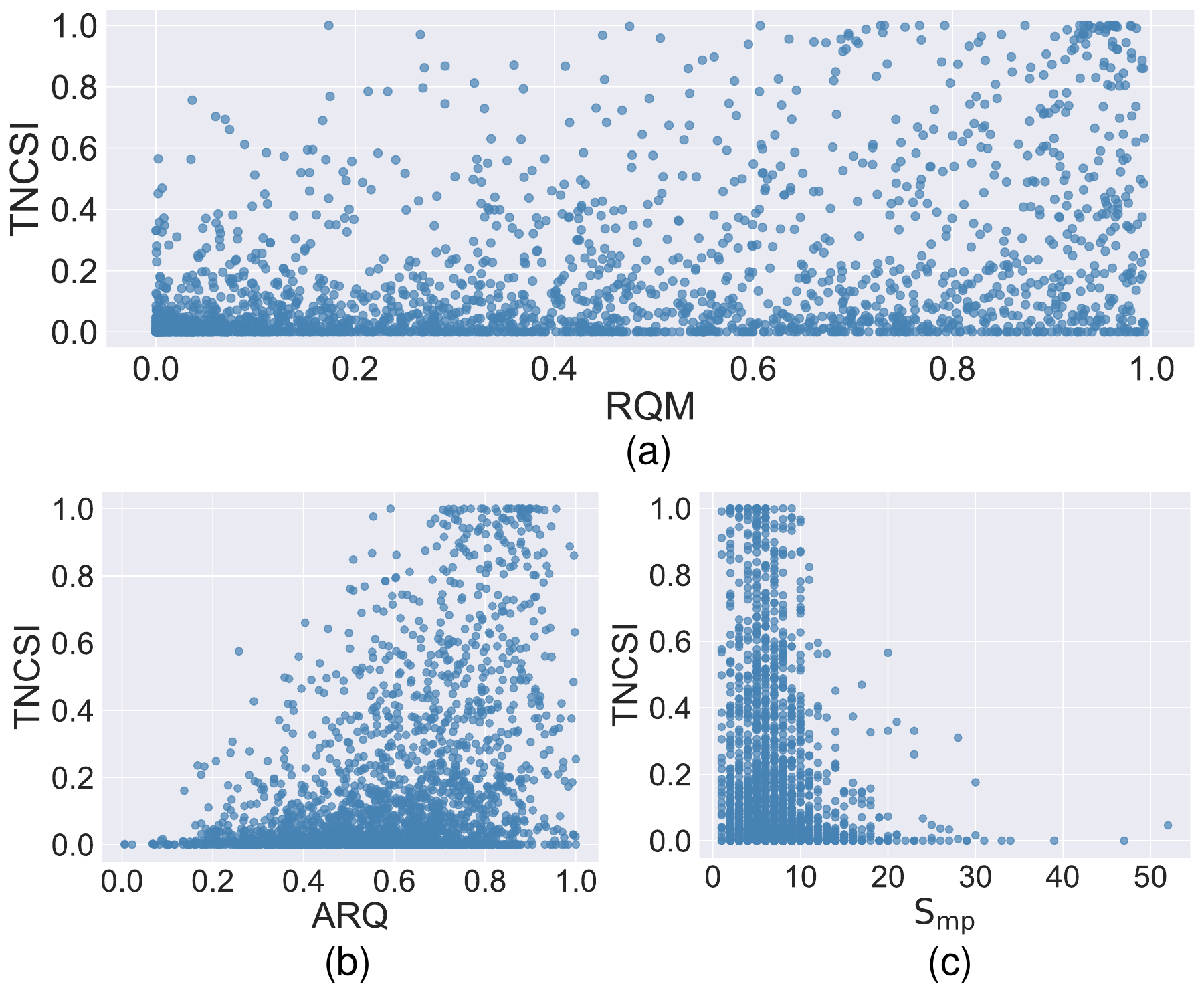}
             \caption{Visualization of the Relationship between Reference Quality and $TNCSI$: Each point corresponds to a review paper. The $TNCSI$ shows positive correlations with $ARQ$ and $RQM$, but a negative correlation with $S_{MP}$.}
             \label{fig:rqm_tncsi}
         \end{figure}

        As the saying goes, \textit{ 'You can't make something out of nothing'}. The same principle applies to literature reviews: a well-crafted review is typically built upon a thorough analysis of high-quality references. To further explore the relationship between the quality of references in literature reviews and the review's impact, we conduct a visual analysis of the distributions of $ARQ$ (representing the quality of references), $S_{MP}$ (representing the recency of references), and the corresponding $TNCSI$ (representing the scholar impact) for reviews published after 2020, as shown in Fig.~\ref{fig:rqm_tncsi}.

        Figure~\ref{fig:rqm_tncsi} (a) highlights the correlation between $RQM$ and $TNCSI$. The scatter plot indicates a notable trend where lower $RQM$ values are generally associated with lower $TNCSI$ values, with a significant concentration of points observed in the range of $TNCSI$ between 0.0 and 0.2. Figure~\ref{fig:rqm_tncsi} (b) demonstrates that reviews citing higher-quality references are more likely to achieve higher academic impact. Generally, it is difficult for reviews with an $ARQ$ below 0.5 to reach a $TNCSI$ higher than 0.5. Furthermore, Fig.~\ref{fig:rqm_tncsi} (c) shows that reviews with a median reference age exceeding 10 semesters (i.e., 5 years) tend to exhibit relatively limited impact.


\begin{table}[ht]
    \renewcommand\arraystretch{1.1}
    \setlength{\tabcolsep}{6pt}
    \centering
    \caption{Correlation between Reference Quality and TNCSI: ``Sig.'' indicates significance, and \checkmark indicates the corresponding p-value less than 0.05.}
    \begin{tabular}{m{0.1\linewidth}m{0.1\linewidth}m{0.21\linewidth}m{0.1\linewidth}m{0.23\linewidth}}
        \hline
        Metric & Pearson & Pearson p-value & Spearman & Spearman p-value \\
        \hline
        \hline
        $ARQ$ & 0.43 & $6.99 \times 10^{-40}$ & 0.42 & $1.11 \times 10^{-37}$ \\
        $S_{mp}$ & -0.27 & $6.29 \times 10^{-16}$ & -0.32 & $2.21 \times 10^{-21}$ \\
        $RQM$ & 0.53 & $1.34 \times 10^{-62}$ & 0.51 & $1.86 \times 10^{-57}$ \\
        \hline
    \end{tabular}
    \label{tab:RQM_corr_analysis}
\end{table}

        To further quantify the relationship between reference features and cumulative academic impact, we present the Pearson and Spearman correlation coefficients between various indicators and $TNCSI$ in Tab.~\ref{tab:RQM_corr_analysis}. Both $ARQ$ and $RQM$ show moderate positive correlations with $TNCSI$, with Pearson and Spearman coefficients around 0.43 and 0.53, respectively. These correlations are highly significant, as indicated by p-values close to zero. In contrast, $S_{mp}$ exhibits a negative correlation, as indicated by Pearson and Spearman coefficients of -0.27 and -0.32, respectively. These statistically validated results demonstrate the importance of reference quality in influencing academic impact of reviews.
        
        \added{Considering that the quality of references exhibits a moderately strong and significant positive correlation with the scholarly impact of a review, $RQM$ can be regarded as an intuitive metric for efficiently gauging the average quality of references. In particular, as observed from Fig.~\ref{fig:rqm_tncsi}, reviews with an $RQM$ greater than 0.8, an $ARQ$ within the range of 0.7 to 0.9, and an $S_{MP}$ not exceeding 10 are more likely to achieve higher academic impact. Illustrative examples of $RQM$ are provided in Tab.~\ref{tab:ex_RQM}.}
        
        \begin{table}[ht]
    \renewcommand\arraystretch{1.1}
    \setlength{\tabcolsep}{6pt}
    \begin{center}
        \caption{Example Use Cases of $RQM$: A higher $RQM$ reflects higher scholarly impact ($ARQ$) and a younger age ($S_{mp}$) of the references.}
        \begin{tabular}{m{0.5\linewidth}m{0.1\linewidth}m{0.1\linewidth}m{0.1\linewidth}}
            \hline
            Title & $ARQ$↑ & $S_{mp}$↓ & $RQM$↑ \\
            \hline
            \hline
            A Survey on Video Diffusion Models~\cite{xing2023survey} & 0.72 & 2 & 0.94 \\
            \hline
            Video Diffusion Models: A Survey~\cite{melnik2024video} & 0.83 & 3 & 0.95 \\
            \hline
            A Survey on Multimodal Large Language Models~\cite{yin2023survey} & 0.83 & 1 & 0.99 \\
            \hline
            Multimodal Large Language Models: A Survey~\cite{wu2023multimodal} & 0.69 & 5 & 0.65 \\
            \hline
            A Survey on Benchmarks of Multimodal Large Language Models~\cite{li2024survey} & 0.52 & 2 & 0.85 \\
            \hline
            A Survey on Evaluation of Multimodal Large Language Models~\cite{huang2024survey} & 0.19 & 2 & 0.63 \\
            \hline
        \end{tabular}
        \label{tab:ex_RQM}
    \end{center}
\end{table}


        \subsection{Tracking Citation Momentum with IEI}
        Although $TNCSI$ enables cross-domain comparisons, it fails to track changes in citation trends. In such cases, the $IEI$ can offer additional insight. A higher $IEI$ suggests that the paper's rate of gaining new citations has been increasing over time, reflecting a growing level of recognition within the academic community. Conversely, an $IEI$ below zero implies that the paper’s citation rate has been declining recently. 
        Table~\ref{tab:ex_IEI} provides examples of $IEI$ applications. In some cases, selecting a paper with a higher $IEI$ may be preferable to choosing one with a higher $TNCSI$, as it reflects a more recent consensus among researchers.
        
        \begin{table}[ht]
    \renewcommand\arraystretch{1.1}
    \setlength{\tabcolsep}{6pt}
    \begin{center}
        \caption{Example Use Cases of $IEI$: A positive $IEI$ indicates that the paper is gaining increasing attention.}
        \begin{tabular}{m{0.5\linewidth}m{0.1\linewidth}m{0.1\linewidth}m{0.1\linewidth}}
            \hline
            Title & Cites & $TNCSI$↑ & $IEI$↑ \\
            \hline
            \hline
            Knowledge Distillation and Student-Teacher Learning for Visual Intelligence: A Review and New Outlooks~\cite{wang2021knowledge} & 565 & 0.99 & -1.41 \\
            \hline
            A Survey on Knowledge Distillation of Large Language Models~\cite{xu2024survey} & 40 & 0.30 & 1.10 \\
            \hline
            A Review of Practical AI for Remote Sensing in Earth Sciences~\cite{janga2023review} & 29 & 0.11 & -0.16 \\
            \hline
            Change Detection Methods for Remote Sensing in the Last Decade: A Comprehensive Review~\cite{cheng2024change} & 23 & 0.23 & 0.40 \\
            \hline
        \end{tabular}
        \label{tab:ex_IEI}
    \end{center}
\end{table}

        \subsection{Assessing the Need for Updating Reviews}

        Literature reviews inevitably age, yet the rate of aging varies significantly across fields. The $RUI$ allows us to gauge the degree of aging of a review. Typically, fields with higher publication rates undergo faster advancements (indicated by a higher $RUI$), meaning that reviews in these areas tend to age more quickly. Conversely, in more established fields with a lower $RUI$, the aging process is correspondingly slower. Compared to relying solely on publication time to assess recency, $RUI$ provides a more accurate estimation. Examples illustrating the exploitation of the $RUI$ are provided in Tab.~\ref{tab:ex_RUI}.
        
        \begin{table}[ht]
    \renewcommand\arraystretch{1.1}
    \setlength{\tabcolsep}{6pt}
    \begin{center}
        \caption{Example Use Cases of $RUI$: $RUI$ quantifies the urgency of updating reviews across different fields. }
        \begin{tabular}{m{0.4\linewidth}m{0.1\linewidth}m{0.2\linewidth}m{0.1\linewidth}}
            \hline
            Title & Year & Topic & $RUI$↓ \\
            \hline
            \hline
            A Review of Object Detection Based on Deep Learning~\cite{xiao2020review} & 2020 & Object Detection & 24.10 \\
            \hline
            A Comprehensive Review of Object Detection with Deep Learning~\cite{kaur2023comprehensive} & 2023 & Object Detection & 6.72 \\
            \hline
            Time-Series Forecasting with Deep Learning: A Survey~\cite{lim2021time} & 2020 & Time-Series Forecasting & 29.09 \\
            \hline
            Long Sequence Time-Series Forecasting with Deep Learning: A Survey~\cite{chen2023long} & 2023 & Time-Series Forecasting & 6.44 \\
            \hline
        \end{tabular}
        \label{tab:ex_RUI}
    \end{center}
\end{table}

        \subsection{Insights into Field Trends via IEI and RUI}
        
\begin{figure}[h]
             \centering
             \includegraphics[width=1.0\linewidth]{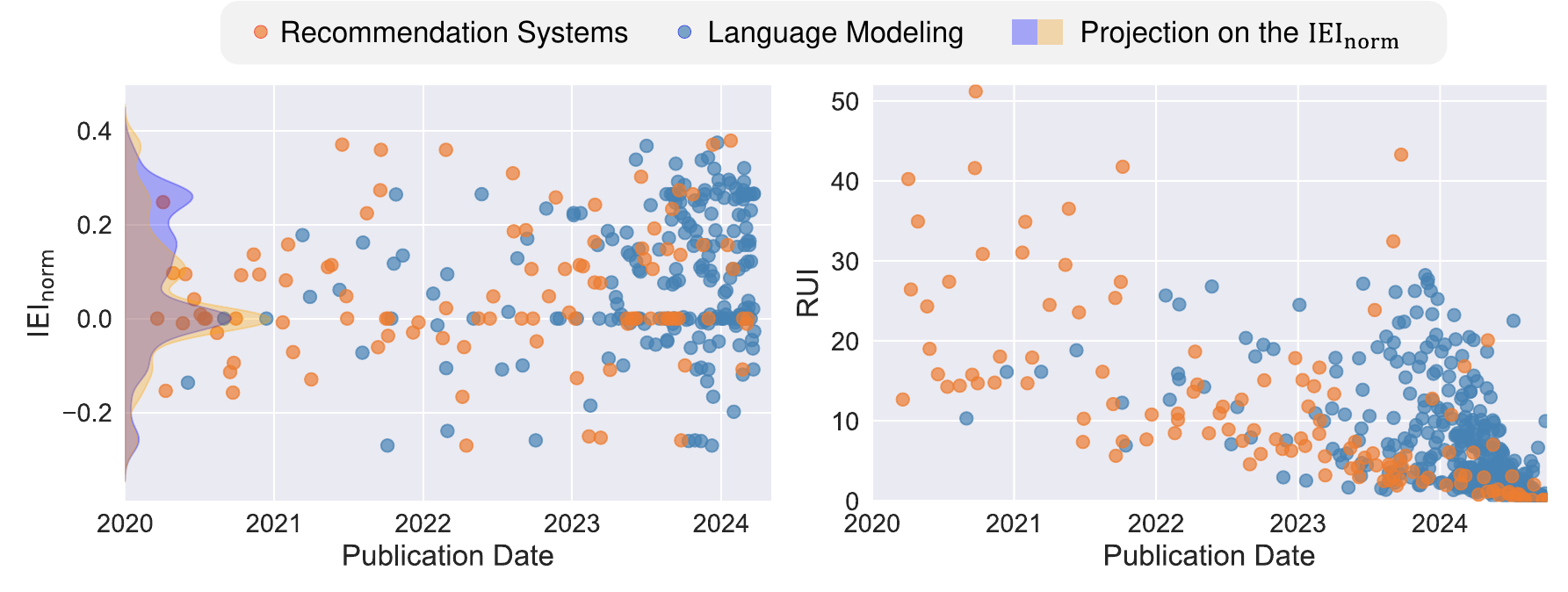}
             \caption{Temporal Distribution of $IEI$ and $RUI$ in Surveys for Recommender Systems and Language Modeling: Higher $IEI$ and $RUI$ values in language modeling over the past two years indicate a trend of intensified development. Although surveys in popular fields tend to attract more academic attention, they also require authors to make greater efforts to keep the content up-to-date.}
             \label{fig:IEI_RUI}
        \end{figure}

        As outlined in Sec.~\ref{sec_IEI} and~\ref{sec_RUI}, $IEI$ and $RUI$ are indicators used to evaluate changes in citation trends and the recency of individual articles. By utilizing these two indicators, we generate scatter plots depicting the temporal distribution of $IEI$ and $RUI$ values for review papers within the domains of recommendation systems and language modeling (as illustrated in Fig.~\ref{fig:IEI_RUI}).
        
        \added{In the left panel of Fig.~\ref{fig:IEI_RUI}, we observe a rapid growth trend in the development of surveys within the field of ``language modeling'' since 2023, which is reflected by the increasing number of surveys with positive $IEI$ values. The right panel of Fig.~\ref{fig:IEI_RUI} further supports this finding. Since 2023, the median $RUI$ for language modeling surveys has noticeably exceeded that of surveys in the recommendation systems field.} As $RUI$ is positively correlated with the volume of publications within a field, a higher $RUI$ suggests a larger publication volume. While this trend underscores the swift progress in the language modeling domain, it also highlights the growing burden on authors to produce and maintain high-quality surveys in such a rapidly evolving and highly competitive area.

        By analyzing the statistical characteristics of numerous review articles' IEI and RUI, we can uncover developmental patterns across various fields. These patterns provide valuable insights for gauging the momentum of progress within each field and advancing exploration in emerging areas.

        \subsection{Integrated Use of the Proposed Indicators}

        In general, researchers are encouraged to apply a set of metrics, rather than relying solely on citation counts to select review articles. In practice, reviews with a $TNCSI$ close to 1 are often prioritized, as this suggests that the review is cited more frequently than most other papers in the same field, indicating widespread esteem within the field. However, supplementary indicators can provide further insights: the $IEI$ captures shifts in citation trends, the $RQM$ evaluates the quality of references, and the $RUI$ reflects how recently the review may require updating. Leveraging a diverse set of indicators allows for the efficient filtering of numerous similar reviews while mitigating the limitations inherent in single-metric citation reliance. 

        Although the proposed metrics provide an intuitive numerical indication, it is important to emphasize that no existing metric, including the proposed ones, can truly capture the intrinsic value of a review (particularly for systematic review). The proposed impact or quality indicators should be regarded as supplementary instruments rather than definitive measures of quality. It is our contention that the true value of a review resides in its capacity to address the specific research questions and meet the needs of its intended audience. 
        More examples of integrated indicators usage and further explanations regarding the fair use of metrics are detailed in the~\ref{sec:appendix_indicators}.
        %
        


    \section{AI-generated Literature Reviews}
    \label{sec_humanvsai}
        \subsection{Overview of AI-generated Literature Reviews}
        Traditionally, literature reviews have been manually conducted by researchers who analyze and synthesize scholarly sources to provide an overview of the current state of knowledge in a specific field. Recently, with the advancement of AI technologies, there's been a growing interest in leveraging artificial intelligence techniques, especially LLMs, to automate or assist in the generation of the literature review~\cite{de2023artificial, tian2024SLSG, Wang2024AutoSurveyLL, lai2024instruct, hu2024hireview}. Typically, users are simply required to indicate their area of research interest, and the system will then automatically generate a literature review.

        The automated creation of a literature review is a multidisciplinary endeavor that integrates knowledge from various fields. It relies not only on artificial intelligence technologies but also requires the merging of knowledge from other fields such as data science, bibliometrics, database engineering, etc. These components are instrumental in extracting, storing, and synthesizing relevant information from vast amounts of literature. Early attempts in AI-generated literature reviews involve training language models, \textit{e.g.}, LLaMA~\cite{touvron2023llama}, on a large corpus of academic papers, research articles, and other scholarly content. These models can then be used to generate coherent and contextually relevant text based on prompts or queries related to a specific research topic. However, given that a fully trained or fine-tuned language model has no access to the latest scholarly advances, these models would not generate a literature review that includes the latest scholarly materials. \added{To overcome this limitation, most advanced AI-generated literature review systems adopt a structured pipeline consisting of four main steps: intention analysis, knowledge retrieval, synthesis, and report generation. The generation of a literature review begins with intention analysis, where the system identifies the research goals and scope. On this basis, the core process consists of knowledge retrieval and synthesis, which are closely interwoven rather than strictly sequential. Retrieval broadens the evidence base by applying criteria such as keywords, publication dates, and citation counts, and by collecting materials not only from academic publications but also from sources such as blogs, tutorials, and open-source repositories. Synthesis complements this step by analyzing the collected resources, identifying relationships among them, and organizing the information into coherent themes. By reinforcing each other, retrieval and synthesis ensure that the system produces a comprehensive and well-grounded review. Finally, the process culminates in report generation, where the integrated content is transformed into a structured and readable document.}

        \added{Beyond their technical implementation, AI-generated literature reviews are often viewed as a promising tool for reducing researchers’ time and effort. However, concerns persist regarding both the reliability and the ethical implications of such content. Zybaczynska~\cite{zybaczynska2023artificial} observes that current AI systems frequently fail to deliver substantial, accurate information or critical judgment. Similarly, Elali \emph{et al.}~\cite{elali2023ai} highlight the risks of fabrication and falsification in AI-generated research, emphasizing their potentially serious consequences for the scientific community. Despite these challenges, there have been notable demonstrations of AI-assisted reviews reaching publication. For instance, Aydın \emph{et al.} used large language models such as ChatGPT and Google Bard to generate reviews in areas including digital twins in healthcare~\cite{aydin2022openai} and the metaverse~\cite{aydin2023google}. Even so, many academic publishers continue to adopt a cautious stance. Leading journals such as \textit{Nature} and \textit{Science}, for example, explicitly prohibit the listing of ChatGPT or other automated tools as paper authors. We contend, nevertheless, that AI-generated reviews—when not intended for direct publication—can still provide substantial value. In particular, they may serve as an efficient aid for researchers seeking to keep pace with rapid developments in fast-evolving fields.}


        \subsection{Current State of AI-generated Literature Review}
        In this subsection, our investigation delves into various efforts that have been made in the field of employing AI techniques to generate literature reviews. 

        \subsubsection{Comparison of Recent AI-generated Literature Review Systems}

\begin{figure}
             \centering
             \includegraphics[width=1.0\linewidth]{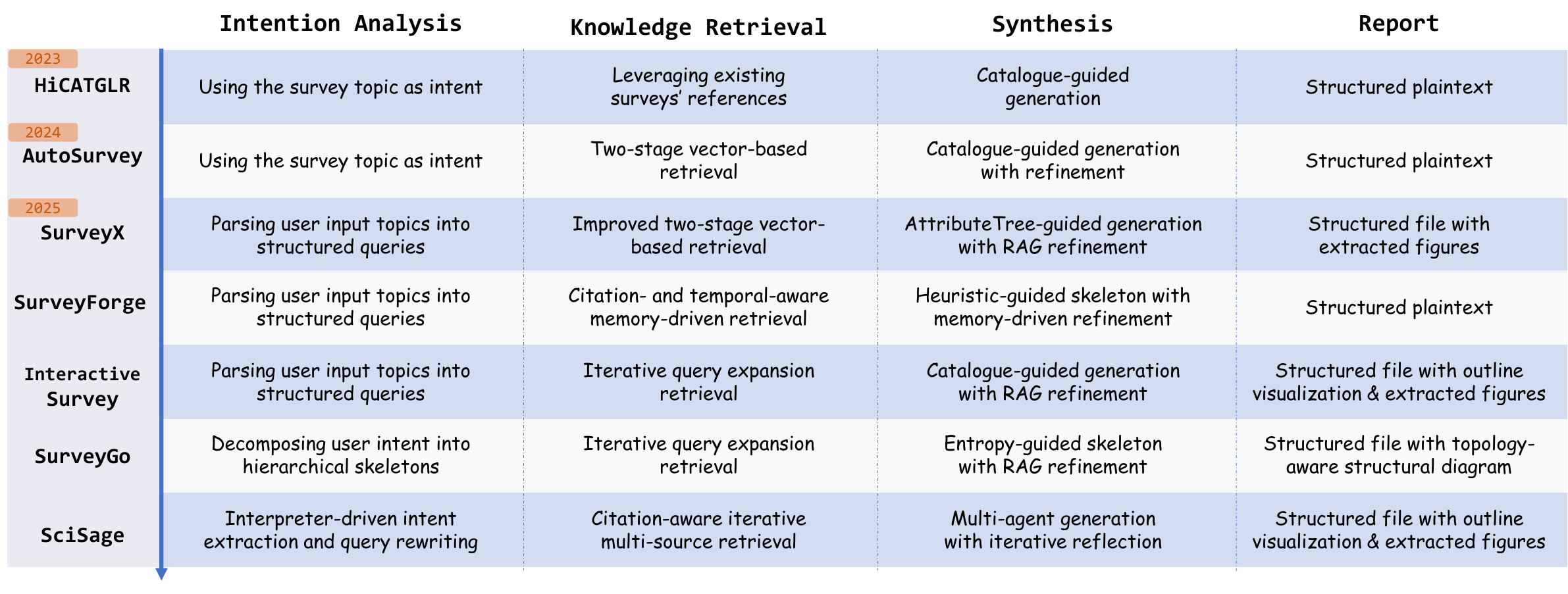}
             \caption{\added{Comparative Overview of Recent Survey Generation Systems: The figure contrasts representative methods (HiCATGLR~\cite{zhu2023hierarchical}, AutoSurvey~\cite{Wang2024AutoSurveyLL}, SurveyX~\cite{liang2025surveyx}, SurveyForge~\cite{yan2025surveyforge}, InteractiveSurvey~\cite{wen2025interactivesurvey}, SurveyGo~\cite{wang2025llm}, SciSage~\cite{shi2025scisage}) across four key stages: intention analysis, knowledge retrieval, synthesis, and report generation. The timeline highlights their evolution from 2023 to 2025.}}
             \label{fig:ai_review_desc}
         \end{figure}

        \added{Positioned as a significant direction in \textit{AI for Research}, automated review generation systems have garnered substantial attention across the research and technology communities. Early explorations—such as Paper Digest~\cite{paperdigest}, Jenni AI~\cite{jenniai}, and various GPT-store plugins~\cite{Seamlees,askyourpdf}—highlight the feasibility of leveraging large language models to generate literature reviews with minimal user input. Building upon this foundation, recent efforts (see in Fig~\ref{fig:ai_review_desc}) have introduced more systematic and scalable approaches that emphasize structured pipelines, citation fidelity, and quality-controlled synthesis. For instance, hierarchical catalogue generation for literature review foregrounds outline construction as a first-class task, releasing a large-scale dataset and semantics-plus-structure metrics that supply supervision for catalogue-guided synthesis~\cite{zhu2023hierarchical}. Building on this foundation, AutoSurvey~\cite{Wang2024AutoSurveyLL} pioneers an end-to-end pipeline—seed retrieval and outline planning, subsection drafting by specialized LLMs, global integration, and iterative evaluation—demonstrating feasibility at scale. To scale capacity and preserve citation fidelity, SurveyGO~\cite{wang2025llm} employs the LLM×MapReduce-V2 “convolutional stacking” architecture and uses an internal benchmark to ensure structural rigor and precise citations. Focusing on workflow quality, SurveyX~\cite{liang2025surveyx} adopts a two-phase Preparation→Generation design that blends offline/online retrieval with an AttributeTree pre-processing pipeline and a re-polishing stage, yielding measurable gains in content quality and citation accuracy. SurveyForge~\cite{yan2025surveyforge} narrows the human–AI gap via heuristic-guided outline construction and a memory-driven scholar navigation agent (SANA), coupled with temporal- and citation-aware re-ranking to improve reference reliability and downstream drafting. Emphasizing user control and multimodality, InteractiveSurvey~\cite{wen2025interactivesurvey} supports outline- and section-wise RAG synthesis and produces reports enriched with structural diagrams and extracted figures/tables. From a reliability perspective, SciSage~\cite{shi2025scisage} introduces a citation-aware multi-agent framework that “reflects while writing,” critiquing drafts at outline, section, and document levels and re-ranking multi-source retrieval by recency and citation signals to strengthen coherence and references.}
    
       \added{
        It is not difficult to observe that, as technologies evolve over time, automated survey systems have gradually matured across the four core components of intention analysis, knowledge retrieval, synthesis, and report generation. Compared to early explorations, many recent systems now produce reviews that are markedly more coherent, informative, and credible. The generated outputs exhibit more reasonable structural organization, richer visual elements such as extracted or synthesized figures and diagrams, and more accurate and verifiable citations. These improvements collectively enhance the readability and reliability of AI-generated reviews, allowing such systems to move beyond proof-of-concept prototypes and begin offering practical value in assisting researchers with staying abreast of emerging fields.
        }

        \subsubsection{Human-authored vs. AI-generated Reviews}
         To highlight the differences in underlying design principles, we provide a descriptive comparison between human-authored and AI-generated reviews. Table~\ref{tab:human_vs_ai} shows that nearly all automatic summarization systems fail to incorporate visual elements. Furthermore, most AI-generated review systems do not include explicit appraisal criteria, making it difficult for users to discern how quality control is enforced and thereby reinforcing the perception of these systems as black boxes.

\begin{table}[!htbp]
    \renewcommand\arraystretch{1.2}
    \setlength{\tabcolsep}{4pt}
    \caption{Comparison Between Human-authored and AI-generated Literature Reviews: "V.E." indicates the presence of visual elements. }
    \begin{center}
        \begin{tabular}{m{0.16\linewidth}  m{0.13\linewidth} m{0.05\linewidth} m{0.55\linewidth}}
            \hline
            Review  & Auto-Level & V.E. & SALSA Analysis~\cite{grant2009typology} \\
            \hline
            \hline
            Human~\cite{wang2023review}  & Manullay & \checkmark & A typical narrative review aims to offer valuable insights in visual prompt learning. The selection criteria for references are not specified, but it's clear that each reference has been thoroughly appraised. Conducts an in-depth synthesis and offers insightful analysis. \\
            \hline
            Jenni~\cite{jenniai}  & Semi-automated &  $\times$ & Automated searching and appraising references are not supported. The generation process relies on user interaction, and only a narrative description is provided. Provides little to no synthesis or analysis. \\
            \hline
            ChatPaper~\cite{kaixindelele_ChatPaper}  & Automated & $\times$ & Retrieving relevant literature from arXiv based on multiple LLM-generated keywords without appraisal step. Each section contains plain descriptions of the related reference. Provides little to no synthesis or analysis. \\
            \hline
            PaperDigest~\cite{paperdigest}  & Automated & $\times$ & Searching papers with the user-specified keyword. It seems to appraise the quality of references with private criteria. The generated content is more like a concise summary. Provides little to no synthesis or analysis. \\
            \hline
            askyourpdf~\cite{askyourpdf}  & Automated & $\times$ &  No official explanations are found for how to retrieve references and appraise the quality of the literature. The generated review includes a brief analysis and description of the related concept, current state of development, and gaps. Provides little to no synthesis or analysis.\\
            \hline
            AutoSurvey~\cite{Wang2024AutoSurveyLL} & Automated & $\times$ &  Narrative review without any visual elements.  Although no formal appraisal framework is utilized, the approach incorporates semantic verification and multi-LLM evaluation. Generated reviews include narrative descriptions of key contributions, research trends, and gaps. Provides little to no synthesis or analysis.\\
            \hline
            \added{SurveyGo}~\cite{wang2025llm}   & Automated & \checkmark & No official explanations are found for how to retrieve references and appraise the quality of the literature. The generated review exhibits a clear outline structure. Offers limited synthesis and analysis.\\
            \hline
        \end{tabular}
    \end{center}

    \label{tab:human_vs_ai}
\end{table}

\begin{figure}
             \centering
             \includegraphics[width=0.7\linewidth]{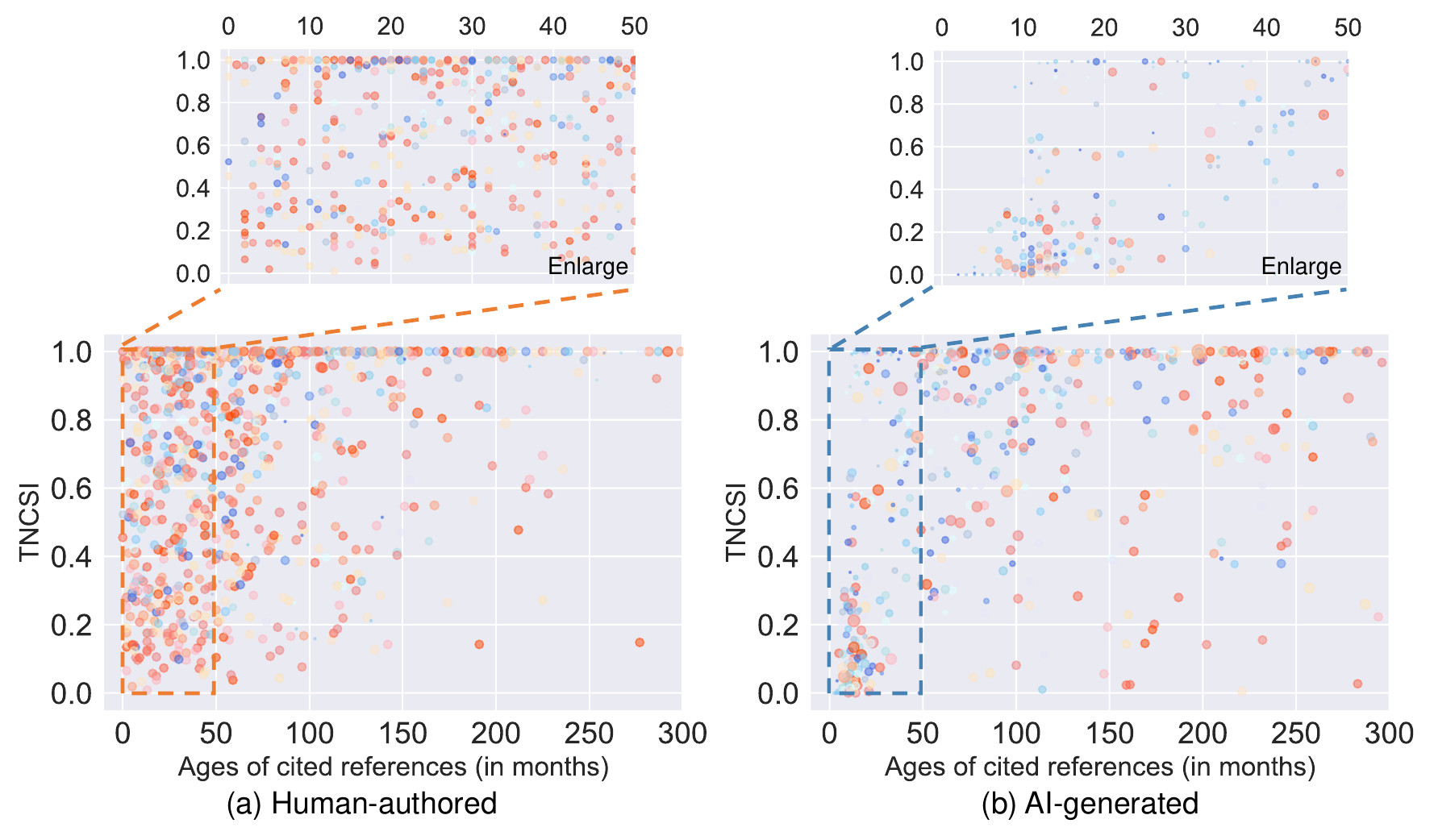}
             \caption{Visualization of References Quality of Human-authored and AI-Generated Literature Reviews: Panel (a) illustrates the quality-age distribution of references in human-written reviews, while panel (b) depicts the quality-age distribution of references in AI-generated reviews. It can be observed that the TNCSI of references less than one year old in AI-generated reviews is significantly lower than that in human-authored reviews. This suggests that current AI-generated review systems struggle to appraise the academic value based on the article content.}
             \label{fig:ref_visualization}
         \end{figure}

        In Fig.~\ref{fig:ref_visualization}, we further present a set of scatter plots depicting the references' quality of (a) reviews published in several academic journals and (b) reviews generated by various AI systems. To allow sufficient time for the references selected by the AI-generated review system to accumulate academic impact, a two-month interval is set between obtaining the reference list and calculating their $TNCSI$. The horizontal axis represents the reference age, and the vertical axis corresponds to the proposed $TNCSI$. The color and size of the scatter points indicate the $IEI$ of the references. A positive $IEI$ value signifies a gradually increasing citation trend, resulting in warm-colored scatter points. The larger the size of the scatter point, the greater the value of $IEI$. Conversely, when the $IEI$ value is negative, the scatter points are cool-colored, and the size decreases as the value decreases. A closer examination of Fig.~\ref{fig:ref_visualization} reveals a distinct distribution among human-authored and AI-generated depicted reviews in their respective scatter plots. Human-authored reviews exhibit clustered dots in the upper left corner, indicating a trend toward referencing more recent and influential sources (\textit{i.e.}, citing newly published works with limited influence at the time of writing). Conversely, AI systems predominantly cite established, high-impact papers, creating a relative sparsity in the upper left corner. We argue that this discrepancy stems from seasoned researchers' ability to assess a paper's value through a diverse, fine-grained, and content-based manner, rather than relying solely on citation counts. This allows them to render more precise judgments about the potential significance of a paper. In contrast, most existing AI-based review systems depend exclusively on citation counts as their primary filtering metric, limiting their capacity to identify high-quality recent publications as references. We also note the presence of a potential boundary line for the $IEI$ metric in Fig.~\ref{fig:ref_visualization}, panel (b), which is likely a result of the quality control mechanisms in certain automated review systems.
        
      \added{Overall, despite continuous technological advances, AI-generated review systems still face several critical limitations. Most systems continue to rely primarily on similarity-based retrieval, with limited attention to the quality of the retrieved literature. Although citation counts are sometimes used as a proxy, they are domain-dependent, temporally biased, and systematically disadvantage newly published work, making them an imperfect signal at best. In addition, current AI systems remain underdeveloped in terms of personalization: they seldom elicit or model users' specific research questions, and thus struggle to generate tailored surveys or guide users toward customized choices. Instead, they often imitate human-written surveys that prioritize coverage, sacrificing the customization that AI could uniquely provide. Even when visual elements are included, they are typically extracted directly from source papers rather than designed as polished, interpretive illustrations (e.g., classification diagrams, timelines, or methodological overviews). Moreover, automated approaches have yet to demonstrate the capability to robustly produce field-level artifacts such as reproducible benchmark tables or domain-specific taxonomies that systematically capture and compare the state of the art. Taken together, these limitations suggest a growing need to explore more reliable, transparent, and intent-aligned approaches to AI-generated literature reviews.}
      
        \subsection{\added{Toward Reliable, Transparent, and Intent-Aligned AI-generated Reviews}}


        \added{Although the capabilities of AI-generated review systems are steadily improving, their adoption in practice remains modest. This may be partly due to ongoing challenges related to reliability, transparency, and the lack of intent-aware customization.}
        
        \added{A key factor limiting the reliability of current systems lies in the lack of mechanisms for evaluating the credibility and quality of referenced literature. Many models retrieve papers based solely on lexical or embedding similarity, without regard to publication venue, peer-review status, or citation history—signals that are often crucial in scholarly contexts. Enhancing reliability may require the integration of explicit citation-quality management components that filter or re-rank literature based on such criteria~\cite{de2024can,zhao2025words}. Furthermore, deeper exploitation of LLMs’ reasoning capabilities~\cite{guo2025deepseek} could allow for more substantive synthesis across sources, moving beyond surface-level summaries toward analytical, comparative, and even critical review writing. Emerging techniques, such as diffusion-based scientific visualization~\cite{chang2025sridbench}, may also contribute to reliability by enabling the automated generation of polished, interpretable visual elements.}
        
        \added{Transparency remains a persistent challenge. Most existing systems function as black boxes: users are seldom informed about how references are retrieved, filtered, or retained. The lack of interpretable intermediate steps or explainable selection criteria undermines trust and limits rigorous evaluation. In contrast, greater transparency in AI-generated reviews can markedly strengthen users’ confidence, as it enables them to better understand and assess the reliability of the outputs. To foster both trust and accountability, AI review systems should disclose their retrieval logic, inclusion and exclusion criteria, and synthesis provenance in a manner that is accessible to both technical and non-technical audiences. Such transparent pipelines not only facilitate critical appraisal but also make it easier to identify errors, biases, and omissions in the generated content.}
        
        \added{Beyond reliability and transparency, one of the most overlooked opportunities of AI-generated reviews may lie in intent-aligned customization. Unlike human-authored reviews—which, once published, must serve a broad and heterogeneous readership—AI systems can adapt outputs dynamically to specific user goals. Depending on the research purpose, a system could generate a concise state-of-the-art overview for practitioners, a mapping review to highlight gaps in the literature, or a full systematic review to support meta-analysis. By tailoring outputs to user intent, such systems can deliver reviews that are more concise, targeted, and practically useful, with flexibility in analytic depth, coverage, and timeliness. This shift toward purpose-driven review generation underscores a unique advantage of AI: not simply replicating established formats, but reimagining the practice of literature reviewing around individual needs and research contexts.}

    \section{Challenges and the Future of the Literature Review in PAMI}
    \label{sec_Challenges_Future}
        Challenges and future opportunities for both human-authored and AI-generated reviews are discussed in this section.
        \subsection{Challenges}
        
        \textbf{Information Overload}.
        While our proposed indicators help alleviate information overload to some extent, several challenges remain unresolved. First, As scientific research continues to expand rapidly, the volume of literature review is expected to grow significantly. Collecting and screening a vast number of publications in these emerging fields will likely lead to incomplete searches. Both human authors and AI systems are required to address the challenge of identifying the most relevant references from the extensive body of published literature. In addition to incomplete searches, the research community should consider exploring ways to avoid redundant efforts. For example, the first two review papers~\cite{zhang2023comprehensive,zhang2023sam} on the segment anything model~\cite{kirillov2023segment}, uploaded to arXiv just two days apart, shared at least 70 overlapping references. Avoiding such duplication not only helps reduce the waste of resources (despite both papers offering unique insights) but also minimizes information redundancy.

        \textbf{Lag and Obsolescence}.
        Scientific knowledge is constantly evolving, yet the process of compiling reviews often struggles to keep up with the latest research advancements—particularly when created by human authors and subject to peer review. Comprehensive reviews are often delayed relative to the emergence of new research areas. For instance, when Mamba~\cite{gu2023mamba} started gaining traction, no in-depth reviews were available during its initial months. This delay reflects the time-intensive nature of gathering, evaluating, and synthesizing an ever-expanding body of literature, along with the requirements of the peer-review process.
  
        \subsection{Future}
        \textbf{Long-term Support Literature Review}.
         The concept of long-term support (LTS) for literature reviews offers a practical solution for maintaining relevance and timeliness in the fast-paced environment of academic research. Unlike traditional reviews, which are generally published once and remain static, an LTS literature review would involve regular updates to include new research findings and advancements in the field. This approach is especially valuable when structured frameworks or taxonomies have been developed, as they serve as foundational tools in their domain and require continuous refinement to remain useful. \added{While such updates may be curated by the original authors to preserve rigor and intent, automated or AI-assisted mechanisms could also play a critical role in monitoring new publications, suggesting revisions, and integrating fresh evidence. In this way, the field benefits from both the continuity of established classifications and the responsiveness afforded by ongoing, partly automated maintenance.}
         
        \textbf{\added{Intent-Aligned and Customized AI-generated Literature Review}}.  
        \added{Human-authored reviews are usually designed to be as comprehensive as possible, aiming not only for broad coverage of existing work but also to serve a heterogeneous readership — from newcomers seeking orientation, to experienced researchers looking for synthesis, and practitioners focusing on applications. In contrast, AI-generated reviews are not constrained by this pursuit of exhaustiveness; instead, they can be tailored to align with specific user intentions and deliver the most relevant literature for given purposes. This ability to generate customized reviews highlights AI’s potential to complement traditional human-authored surveys by providing focused, intent-driven perspectives. Future development may further enhance this customization by integrating precise content-based evaluations and visual elements, thereby improving both the relevance and readability of AI-generated reviews.}

\section{Conclusion}
\label{sec_conclusion}

\added{This study provides the first large-scale tertiary analysis of literature reviews in the PAMI field and yields several key conclusions.} 

\added{First, our structural investigation of more than 3,000 reviews uncovers highly consistent organizational anchors—such as the placement of introductions and conclusions—yet also reveals systematic variation in and limited adoption of methodological reporting standards. Features like benchmarking, preliminaries, and structured abstracts are gaining momentum, but overall practice remains uneven across subfields and publication types. These findings underscore both the strengths and the blind spots of current review-writing conventions.}

\added{Second, our proposed bibliometric indicators ($TNCSI$, $RQM$, $IEI$, and $RUI$) demonstrate concrete value in navigating the rapidly growing body of surveys. They help normalize impact across domains, highlight the role of reference quality, track citation momentum, and identify the timeliness of reviews. Together, these measures provide researchers with practical tools for screening overlapping surveys and reducing over-reliance on raw citation counts. They also supply principled signals for future AI-for-Research systems that may incorporate human-authored reviews as seeds for automated synthesis.}

\added{Third, our comparative assessment of AI-generated and human-authored reviews points to both progress and limitations. While recent AI systems show improved coherence, organization, and even integration of visual elements, they remain constrained by over-reliance on citation counts, weak personalization, and difficulty in capturing very recent advances. Human experts still excel at nuanced appraisal of new contributions, but AI systems hold promise as intent-driven complements that can deliver customized perspectives for different user needs.}

\added{We release our code framework for metadata retrieval, indicator computation, and statistical analysis. Although designed for PAMI, the approach is readily extensible to other domains, and we hope it will serve as a foundation for more reliable, transparent, and purpose-driven review practices across research communities.}

\section*{Acknowledgements}
This research was supported by the Fund of the National Natural Science Foundation of China (Grant No.62576177, 62206134), the Fundamental Research Funds for the Central Universities 070-63233084, and the Tianjin Key Laboratory of Visual Computing and Intelligent Perception (VCIP). Computation is supported by the Supercomputing Center of Nankai University (NKSC). This work was supported by the National Science Fund of China under Grant No. 62361166670. ChatGPT is utilized to refine some of the sentences.
%
%
%
\bibliographystyle{splncs04}
\bibliography{egbib}
%

	\newpage
	
\begin{appendix}

\section{Details of the Proposed Indicators.}
\label{sec:appendix_indicators}
        
        In this section, we provide further insights into the proposed indicators. As presented in Tab.~\ref{tab:metric_compare}, various previous efforts have aimed to tackle the challenge of comparing metrics across fields.  For instance, Paper~\cite{purkayastha2019comparison} analyzes two well-known field-independent citation metrics at the article level: Field-Weighted Citation Impact (FWCI) and Relative Citation Ratio (RCR). This study suggests that both metrics perform similarly in normalizing citations across research domains. However, these metrics require predefined field categorizations (such as the Scopus All Science Journal Classification) and may not adequately assess papers in emerging subfields. The Field Normalized Citation Success Index (FNCSI), introduced in Refs.~\cite{shen2020utilization, tong2023novel}, is defined as the likelihood that a paper published in Journal A is cited more frequently than a randomly selected paper from Journal B. Although FNCSI offers robustness, it should be noted that it depends on predefined topic keywords and is specifically suited for journal-level assessments.

        As the Leiden Manifesto~\cite{hicks2015bibliometrics} states: metrics should ``\textit{Account for variation by field in publication and citation practices}''. We develop the concept of impact indicators, which is a measure used to gauge the impact of a certain paper in its field. The reason for using the term ``indicator'' rather than ``metric" is we believe that the impact of a certain paper cannot be fully and accurately measured. While metrics like citation counts, h-index, or journal impact factors could indicate a paper's influence within the academic community, they fail to capture all aspects of its impact. For instance, a research paper might lead to significant advancements in theory, methods, or understanding in its field, none of which would necessarily be reflected in academic metrics. Similarly, a paper might contribute novel concepts or techniques that become influential over time but are not initially evident in citation counts. Hence, the term ``indicator'' is favored, as it suggests a signal without asserting to encompass the entirety of the academic paper's contribution.

   
   
	

\begin{table}[ht]
    \renewcommand\arraystretch{1.1}
    \setlength{\tabcolsep}{6pt}
    \begin{center}
        \begin{tabular}{p{0.23\linewidth}p{0.18\linewidth}p{0.2\linewidth}p{0.20\linewidth}}
            \hline
            Metric & Assessing Level & Normalized & Pre-defined Keywords Free  \\
            \hline
            \hline
              Citation Counts  & A/J & $\times$ & \checkmark \\
            \hline
              Impact Factor  & J & $\times$ & $\times$ \\
            \hline
              FNCSI~\cite{tong2023novel}  & J & Field and Value & $\times$ \\
            \hline
              CiteScore~\cite{teixeira2020citescore}  & J & $\times$ & $\times$ \\
            \hline
              SNIP~\cite{moed2010measuring}  & J & Field & $\times$ \\
            \hline
              FWCI~\cite{FWCI}  & A & Field & $\times$ \\
            \hline
              RCR~\cite{hutchins2016relative}  & A & Field & $\times$ \\
            \hline
              TNCSI(\textit{Ours})  & A/J & Field and Value & \checkmark \\
            \hline
        \end{tabular}
    \end{center}
    \caption{\textbf{Metrics for Evaluating Scholar Impact of Papers:} ``A'' and ``J'' stand for article-level and journal-level. Field normalized signifies that the metric can be utilized across fields. Value normalized indicates that the range of the value is between 0 and 1.}
    \label{tab:metric_compare}
\end{table}

        \subsection{TNCSI}
        \label{sec:appendix_TNCSI}

\begin{figure}
             \centering
             \includegraphics[width=0.5\linewidth]{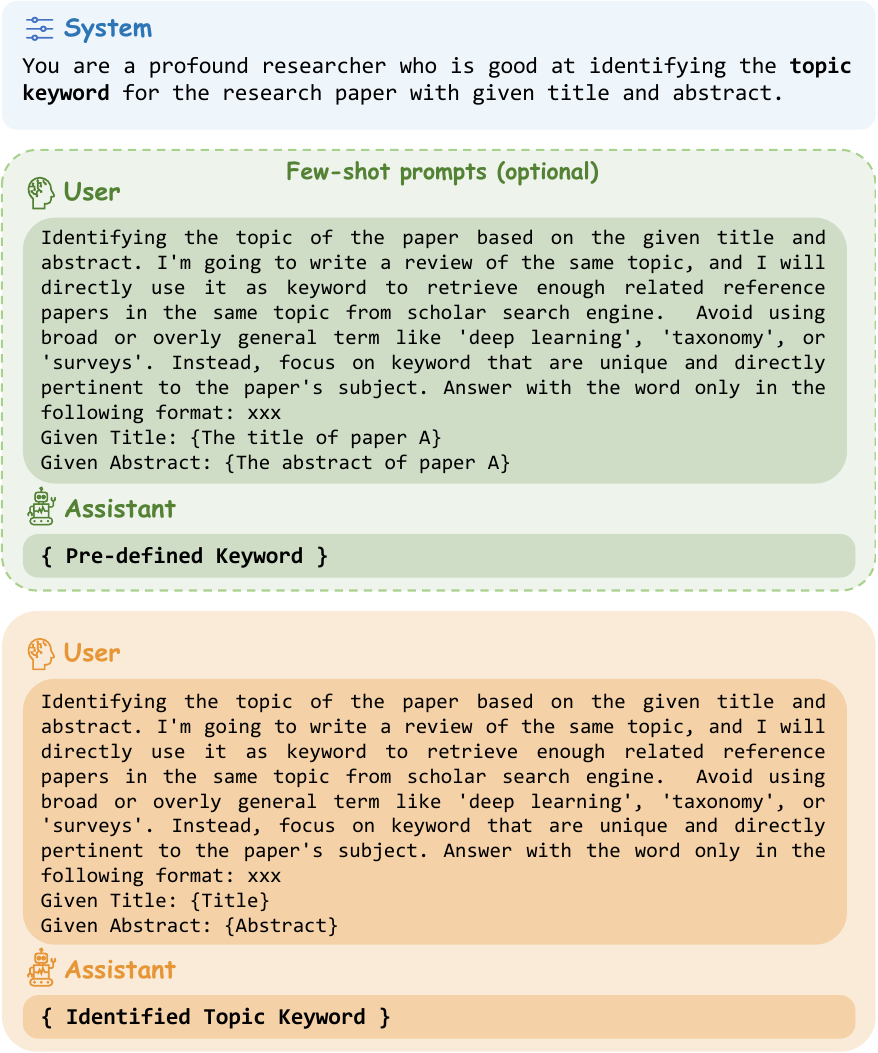}
             \caption{Conceptual Illustration of Topic Keyword Generation Process: Few-shot prompting may enhance the response quality of the large language model.}
             \label{fig:prompt}
        \end{figure}

        $TNCSI$ is a metric used to assess the academic impact of articles across various disciplines. The fundamental distinction of $TNCSI$ from other metrics lies in its independence from the predefined keyword. The requirement for a predefined keyword is a primary reason that most bibliometric indicators face challenges in facilitating cross-disciplinary comparisons. To address such a limitation, we propose adopting ChatGPT~\cite{openai2022chatgpt} (gpt-3.5-turbo-0125) to generate the topic keyword and retrieve related papers according to the keyword. As illustrated in Fig.~\ref{fig:prompt}, we ask ChatGPT to identify the most representative topic keyword with the paper's title and abstract. ChatGPT is one of the most advanced and influential LLMs in the field of natural language processing~\cite{zhao2023survey}. Equipped with state-of-the-art language understanding capabilities, ChatGPT has revolutionized the way we interact with AI-powered conversational systems by simply setting ``system'', ``user'', and ``assistant'' roles. The ``system'' role sets the conversation's behavior and initial context. It provides instructions to guide the assistant's responses. The ``user'' role represents the individual interacting with ChatGPT, who inputs messages to the assistant. The ``assistant'' role is the ChatGPT model itself which would respond based on the provided instructions and user input.

        We count the citations of papers with the help of Semantic Scholar API. For each paper, we retrieved up to $k=1000$ relevant sources using the API. Based on the selected $k$ papers and the corresponding citation counts $p_c$ for each paper $p$, we can calculate the discrete citation frequency distribution of the $k$ papers in a certain topic. We may further consider the  distribution as a probability mass function:
        
        \begin{equation}
        \label{eq_c_k}
            P(X=x) = 
            \begin{array}{ll} 
                \frac{{\text{Citation}_x}}{k},
            \end{array}
        \end{equation}
         where $Citation_x$ represents the number of papers with $x$ citations. The $k$ papers related to the topic and their meta-data could be retrieved using a pre-defined topic keyword through online scholar search engines or API, such as Semantic Scholar, CrossRef, or Google Scholar.
        
        Considering that the number of papers with $x$ citations generally follows an exponential decay, we utilize the maximum likelihood estimation method to fit $P(X=x)$ and obtain the probability density function (PDF) of a continuous exponential decay distribution:
    
        \begin{equation}
        \label{eq_PDF}
            f(x) = \lambda e^{-\lambda x}, x \geq 0.
        \end{equation}
            
        Next, we will calculate $TNCSI$ by performing a definite integral on $f(x)$.
        \begin{equation}
            TNCSI = \int_0^{\text{citeNum}} f(x) \,dx , x \geq 0, 
        \end{equation}
        where $f(x)$ represents the probability density at the value $x$, and $\lambda$ is the results obtained from the maximum likelihood estimation, representing the scale parameter controlling the scaling. Finally, the definite integral of $f(x)$ over the interval $[0, citeNum]$ gives us the desired $TNCSI$:

    In most scenarios, the generated topic word is sufficient to meet expectations, which can be further used as the keyword to retrieve papers from online scholarly search engines. Optionally, one can set ``System'', ``User'', and ``Assistant'' roles before the final query to improve response quality and create more tailored interactions with the ChatGPT. In other words, a few-shot user-assistant pair prompts the ChatGPT with context on topic granularity. For example, a paper about the classification of irises by an improved CNN may have different perspectives. Some researchers focus more on the algorithm of the improved CNN, while others may be interested in classifying irises. Such ambiguity would likewise make it difficult for ChatGPT to identify the most representative topic keyword as expected. However, this could be addressed by providing the context which consists of (1) the identical prompt template with the replaced title and abstract as user input, and (2) the expected topic keyword as assistant output. By default, we adopt a well-known paper~\cite{dosovitskiy2020image} as an example to guide models to generate the topic keyword for all papers.

     Similar to other LLMs, ChatGPT adapts its responses based on user-provided natural language prompts. However, natural language prompts can be ambiguous, and different prompts may lead the model to respond with varying quality. Thus, we follow the practices of LLM prompt engineering and carefully design the prompt to optimize the desired output. To determine the optimal prompt, we construct a dataset by manually annotating the topic keywords of 201 papers from various domains published later than the ChatGPT being trained and then compare the performance of multiple prompts on this dataset. The normalized edit distance~\cite{yujian2007normalized} is adopted to measure the similarity between the GPT-generated keyword and our annotated keyword, where a lower value indicates a higher quality of the prompt. As can be seen from Tab.~\ref{tab_topickwd}, some of the designed prompts achieve decent NED scores for papers in various domains.
    
   
   
	

\begin{table*}[t]
    \renewcommand\arraystretch{1.1}
    \setlength{\tabcolsep}{6pt}
    \begin{center}
        \begin{tabular}{p{0.02\linewidth}p{0.75\linewidth}p{0.08\linewidth}p{0.05\linewidth}}
            \hline
            NO. & User Prompt Content & Few-shot & NED↓ \\
            \hline
            \hline
            1 & Please analyze the title and abstract provided below and identify the main topic or central theme of the review paper. Focus on key term and the overall subject matter to determine the primary area of research or discussion.The output should be formatted as following: xxx  & $\times$  & 0.75 \\
            \hline
            2 & Given title and abstract, please provide the searching key phrase for me so that I can use it as keyword to search highly related papers from Google Scholar or Semantic Scholar. Please avoid responding with overly general keyword such as deep learning, taxonomy, or surveys, etc. Answer with the words only in the following format: xxx  & $\times$  & 0.40 \\
            \hline
            3 & Identifying the topic of the paper based on the given title and abstract. Avoid using broad or overly general term like 'deep learning', 'taxonomy', or 'surveys'. Instead, focus on keyword that is unique and directly pertinent to the paper's subject. Answer with the word only in the following format: xxx & $\times$  & 0.36 \\
            \hline
            4 & Identifying the topic of the paper based on the given title and abstract. So that I can use it as keyword to search highly related papers from Semantic Scholar.  Avoid using broad or overly general term like 'deep learning', 'taxonomy', or 'surveys'. Instead, focus on keyword that is unique and directly pertinent to the paper's subject. Answer with the word only in the following format: xxx   & $\times$  & 0.32 \\
            \hline
            5 & Identifying the topic of the paper based on the given title and abstract. I'm going to write a review of the same topic and I will directly use it as keyword to retrieve enough related reference papers in the same topic from scholar search engine.  Avoid using broad or overly general term like 'deep learning', 'taxonomy', or 'surveys'. Instead, focus on keyword that are unique and directly pertinent to the paper's subject. Answer with the word only in the following format: xxx   & $\times$  &  0.29 \\
            \hline
            6 &  Identifying the topic of the paper based on the given title and abstract. I'm going to write a review of the same topic and I will directly use it as keyword to retrieve enough related reference papers in the same topic from scholar search engine.  Avoid using broad or overly general term like 'deep learning', 'taxonomy', or 'surveys'. Instead, focus on keyword that are unique and directly pertinent to the paper's subject. Answer with the word only in the following format: xxx   & $\checkmark$  &  \textbf{0.28} \\
            \hline

        \end{tabular}
    \end{center}
    \caption{Effectiveness of Prompt Engineering: The few-shot approach slightly improves the performance of topic keyword extraction.}
    \label{tab_topickwd}
\end{table*}

    \added{It is noteworthy that the proposed indicator exhibits a certain degree of robustness with respect to the choice of keywords. This robustness stems from the underlying mechanisms of modern search engines, which rely primarily on semantic similarity rather than strict keyword matching. As a result, even when the keywords differ, the retrieved literature remains largely consistent. In Table~\ref{tab:synonym_entropy}, we report the KL divergences of citation distributions obtained from searches with different but semantically related keywords. The results show that the average KL divergence remains below 0.1, indicating that the metric is highly stable under variations in keyword phrasing.}

    \begin{table}[ht]
    \renewcommand\arraystretch{1.15}
    \setlength{\tabcolsep}{6pt}
    \centering
    \begin{tabular}{p{0.25\textwidth} p{0.5\textwidth} >{\raggedleft\arraybackslash}p{0.15\textwidth}}
        \hline
        Anchor Term & Comparison Terms & Avg. KL \\
        \hline\hline
        Object Detection & Target Detection, Object Localization & 0.080 \\
        \hline
        Face Detection & Facial Landmark Detection, Face Recognition & 0.052 \\
        \hline
        Image Classification & Object Categorization, Visual Classification & 0.086 \\
        \hline
        Pose Estimation & Human Pose Detection, Body Pose Estimation & 0.194 \\
        \hline
        Semantic Segmentation & Scene Segmentation, Class-Level Segmentation & 0.049 \\
        \hline
        Generative Adversarial Networks & GAN, Adversarial Networks & 0.014 \\
        \hline
        Object Tracking & Visual Tracking, Moving Object Detection & 0.075 \\
        \hline
        Image Super-Resolution & Image Enhancement, Super-Resolution Imaging & 0.014 \\
        \hline
        Action Recognition & Activity Recognition, Gesture Recognition & 0.090 \\
        \hline
        3D Reconstruction & Three-Dimensional Reconstruction, Scene Reconstruction & 0.010 \\
        \hline
        Pedestrian Detection & Person Localization, Human Detection & 0.042 \\
        \hline
        Vehicle Detection & Car Detection, Automobile Detection & 0.113 \\
        \hline
        Text-to-Image Generation & Visual Synthesis, Image Generation from Text & 0.020 \\
        \hline
        Neural Style Transfer & Style Transfer, Image Transformation & 0.147 \\
        \hline
        Speech Recognition & Voice Recognition, Speech-to-Text & 0.044 \\
        \hline
        Object Recognition & Object Identification, Visual Recognition & 0.076 \\
        \hline
        Medical Imaging & Radiology Imaging, Clinical Imaging & 0.011 \\
        \hline
        Robotic Vision & Machine Vision, Automated Vision & 0.022 \\
        \hline
        Data Augmentation & Data Synthesis, Synthetic Data Generation & 0.179 \\
        \hline
        Human-Computer Interaction & User Interface Interaction, HCI & 0.005 \\
        \hline
        Scene Graph Generation & Object-Relationship Graph, Contextual Graphs & 0.161 \\
        \hline
        \textbf{Overall} & --- & 0.071 \\
        \hline
    \end{tabular}
    \caption{\added{Average KL Divergence of Synonym Groups: The metric is computed in a single direction with the first term of each group as the anchor, and the ``Comparison Terms'' column lists the alternative expressions. These values are generally low, indicating robustness under keyword variations.}}

    \label{tab:synonym_entropy}
\end{table}

\begin{figure}[H]
         \centering
         \includegraphics[width=0.9\linewidth]{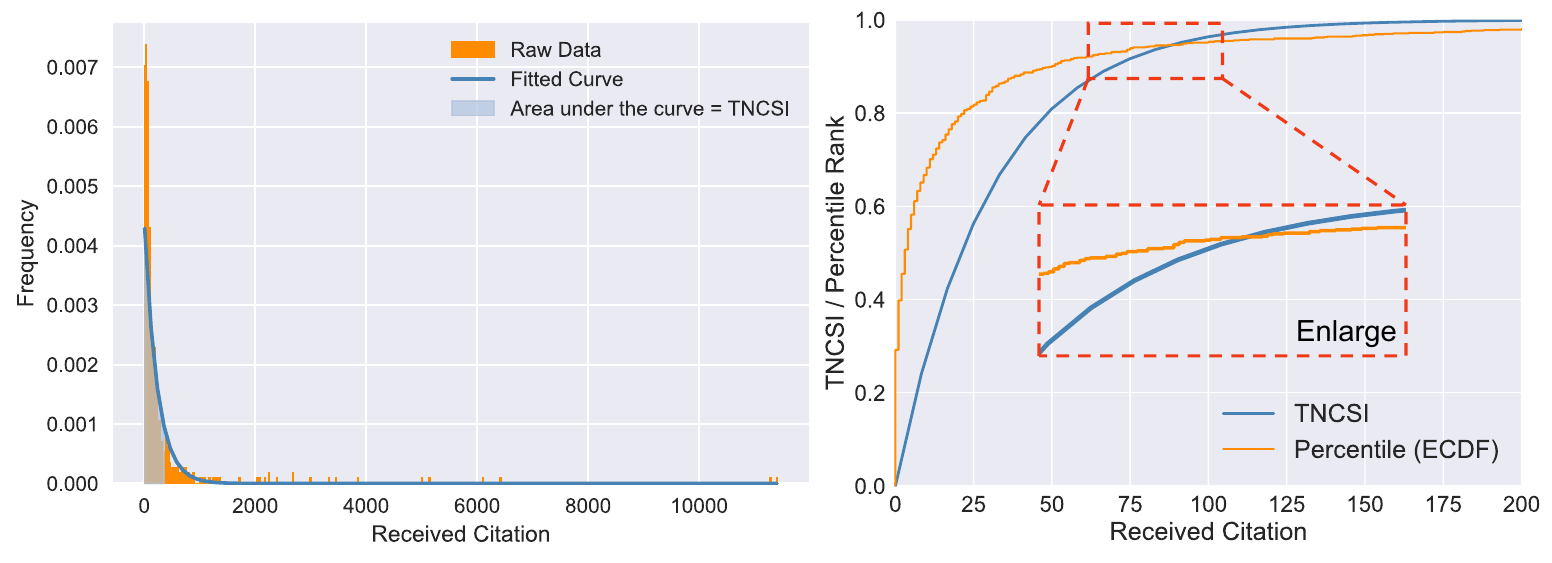}
         \caption{\added{Illustration of the Proposed Quality Indicators: 
            The $TNCSI$ is defined as the area under the fitted probability density function curve estimated via maximum likelihood, resulting in a smooth cumulative representation. 
            In contrast, the percentile-based citation indicator corresponds to the empirical cumulative distribution function (ECDF), which is discrete and may assign the same percentile value to different citation counts.}}

         \label{fig:TNCSI_illustration}
     \end{figure}

    To provide a clear understanding of the proposed quality indicators, we render the graphical representations of $TNCSI$ in Fig.~\ref{fig:TNCSI_illustration} left panel. As can be seen in Fig.~\ref{fig:TNCSI_illustration}, $TNCSI$ equals the area under the probability density function curve, which is fitted with the use of maximum likelihood estimation. \added{In the right panel, $TNCSI$ is contrasted with the percentile-based citation indicator, which corresponds to the empirical cumulative distribution function (ECDF). Unlike the smooth curve of $TNCSI$, the ECDF is discrete, meaning that different citation counts may share the same percentile value. This discreteness can introduce unavoidable errors in certain tasks such as network training~\cite{zhao2025words}.}
    
    \subsection{IEI}
\label{sec:appendix_IEI}

        The $IEI$ is an indicator used to measure changes in citation trends over a given period. A direct method for evaluating changes in citation trends is to apply polynomial fitting to the scatter data and calculate the sum of the derivatives at each point. However, in practice, polynomial fitting methods tend to be sensitive to outliers when applied to data lacking discernible distribution patterns, potentially compromising their robustness significantly. Consequently, the numerical values obtained may not accurately reflect the underlying citation trend. In contrast to these series analysis-based methods, we propose a morphological theory-grounded Impact Evolution Index ($IEI$), which converts the citation trend into a clear and interpretable numerical value.
        
        To calculate the $IEI$, we first need to obtain the distribution of citation counts over time since publication. Once the citation data is retrieved, we may create a sequence $Seq_\text{citation}$ about the number of citations. The $i\in\{0, 1, 2 ..., l\}$ item in the sequence $Seq_\text{citation}[i]$ represents the number of received citations in the $i_\text{th}$ month after the publication, where $l$ represents the number of months allocated for trend observation. Then, a sequence $Seq_\text{time}$ of the same length as $Seq_\text{citation}$ is generated by enumerating from 0 to $l-1$,  Typically, the minimum recommended value for $l$ is 6 or higher. This ensures that the data used for the analysis is adequately representative and the results are reliable. We match the items at the same positions in $Seq_\text{time}$ and $Seq_\text{citation}$ to determine a set of discrete coordinates $\{(Seq_\text{time}[i], Seq_\text{citation}[i])\}$, which serve as the control points for shaping the Bézier curve. A Bézier curve is a mathematical representation of smooth curves commonly used in computer graphics, image editing, and design software. The curve starts at the first control point and ends at the last control point, while the intermediate control points influence the curvature and direction of the curve. The number of control points determines the degree $n=l-1$ of the curve.
        \begin{equation}
            \label{eq_beizer}
            C(t) = \sum_{i=0}^{n} B_{i,n}(t) P_i,
        \end{equation}

        \begin{equation}
            \label{eq_bezier_coefficient}
            B_{i,n}(t) = \binom{n}{i} (1-t)^{n-i} t^i, t \in [0,1],
        \end{equation}
        where $B_{i,n}(t)$ represents the coefficient of the Bézier curve at a given parameter value $t$, which determines the position along the curve ($t=0$ means the start and $t=1$ means the end). $\binom{n}{i}$ is the binomial coefficient, also known as ``n choose i". It represents the number of ways to choose $i$ elements from a set of $n$ elements. $P_i$ stands for the $i_\text{th}$ control point of the curve.

        Given the continuity of the Bézier curve, we can compute its derivative as follows:
        \begin{equation}
            \label{eq_beizer_derivate}
            C'(t)=n\cdot\sum_{i=0}^{n-1}B_{i,n-1}(t)\cdot\left(P_{i+1}-P_{i}\right).
        \end{equation}
        
        The tangent vector $C'_a$ at the $a$-th point on the Bézier curve is further given by Eq.~\eqref{eq_beizer_derivate_vector}. 
        \begin{equation}
        \label{eq_beizer_derivate_vector}
        \begin{aligned}
        C'_a = \ & n\cdot\sum_{i=0}^{n-1}B_{i,n-1}(\frac{a}{n})\cdot\left(P_{i+1}-P_{i}\right), \\
        = \ & (x_a,y_a) , \ a=0,1,\cdots,n,
        \end{aligned}
         \end{equation}
         where $x_a$ and $y_a$ are components of the vector, representing its magnitude along the x and y axes respectively.
         
        Finally, the $IEI_{L_l}$ can be obtained by averaging the slope of $l=n+1$ distinct points on the curve.  Moreover, different months may contribute differently to the $IEI$. For instance, if we desire closer months to have a greater impact, we can achieve this by adjusting the weighting coefficients, $w_a$, of the slope at different points to calculate their weighted averages (See in Eq.~\eqref{eq_IEI_Weight}). In addition, the instantaneous trend could be regarded as the slope of the last month in the sequence. It can be obtained by setting $w_n=1$ and the other weighting coefficients to 0. We denote the $IEI$ focused on the last month among the latest $l$ months (excluding the current month) as $IEI_{I_l}$, as depicted in Eq.~\eqref{eq_IEI_insta}.

         \begin{equation}
        \begin{aligned}
            \label{eq_IEI_Weight}
            IEI_{W_l} = \sum_{a=0}^{n} \frac{w_a (y_a/x_a)}{n+1},
        \end{aligned}
        \end{equation}
        \begin{equation}
            \label{eq_IEI_insta}
            IEI_{I_l} = C'(1) = n\cdot(P_n-P_{n-1}).
        \end{equation}
        Note that the value of $l$ can be configured flexibly to meet actual demands. In general, the longer the period being analyzed, the more stable the citation trend becomes. We usually prefer to analyze the most recent 6 months of citations when constructing a Bézier curve of degree 5 and calculate $IEI_{L_6}$, $IEI_{W_6}$, and $IEI_{I_6}$ based on the curve.

\begin{figure}[H]
         \centering
         \includegraphics[width=1\linewidth]{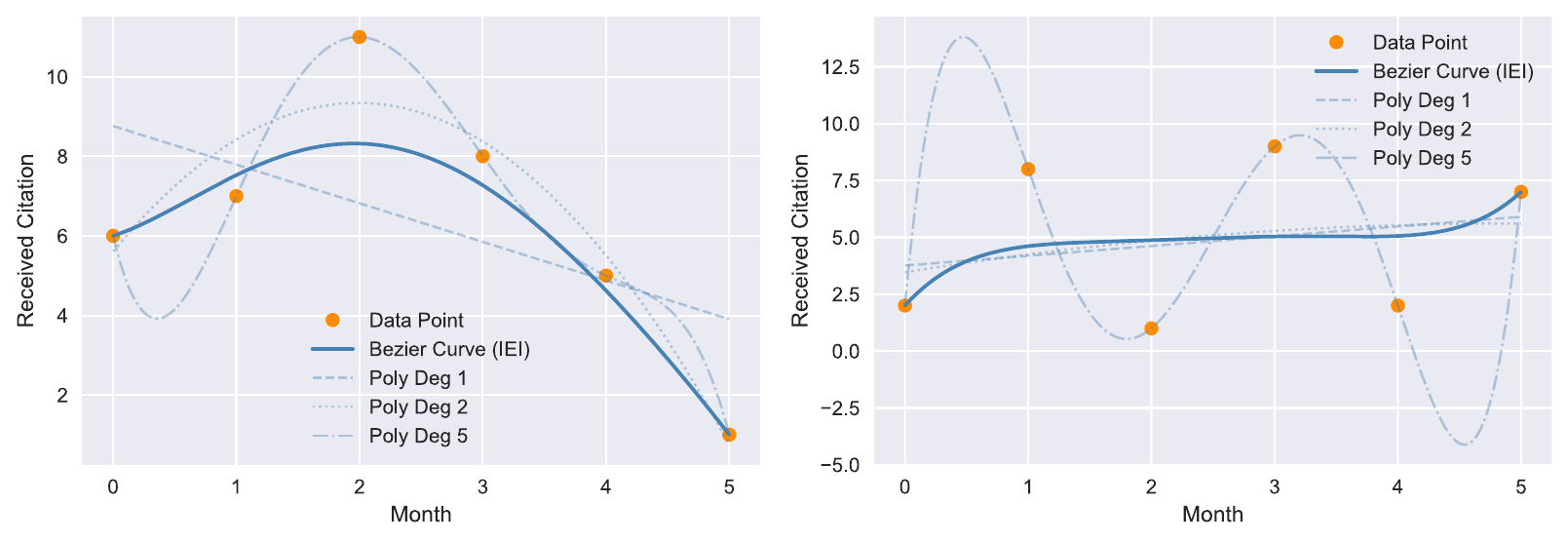}
         \caption{\added{Visualization of the Proposed $IEI$: $IEI$ is the average of the derivatives of each control point on the Bézier curve.}}
         \label{fig:IEI_illustration}
     \end{figure}

        The illustration of the $IEI$ is shown in Fig.~\ref{fig:IEI_illustration}. In Fig.~\ref{fig:IEI_illustration}, the horizontal axis labeled 0 to 5 inversely denotes the months prior to the current month, with 0 representing 6 months ago and 5 denoting the previous month. \added{From the figure it can be observed that, compared with the fifth-degree polynomial fitting, the $IEI$ avoids overfitting and produces a smoother and more stable trend. In contrast to the first- or third-degree polynomial fittings, which either oversimplify the dynamics or introduce spurious inflection points, the $IEI$ achieves a balanced representation that captures the overall trajectory while maintaining robustness against local noise.}

    \subsection{RQM}
    \label{sec:appendix_RQM}
        The quality of references in a literature review is a multifaceted concept involving various aspects, such as credibility, relevance, breadth, and depth. Quantitatively assessing these elements is challenging. To begin with, precisely defining the relevance of scientific literature is difficult; both co-citation analysis~\cite{small1973co} and similarity in paper embeddings have inherent conceptual limitations. For instance, while the concept of breadth in a literature review can theoretically be measured as the ratio of cited references to the total number of relevant references, accurately calculating this ratio is challenging. Identifying all relevant references through keyword searches or citation networks remains elusive.
        
        Given the challenges in quantifying reference quality through the direct factors mentioned, we instead consider an indirect indication of reference quality using $TNCSI$. Here, $TNCSI$ serves as a quality indicator based on collective user assessments, effectively capturing a statistical summary of numerous researchers' evaluations of these direct factors for a given paper. Timeliness also matters. An up-to-date literature review ensures that the latest advancements, developments, and perspectives within a field are incorporated. This temporal relevance enhances the accuracy and effectiveness of research outcomes by reflecting the current state of knowledge. By emphasizing the currency of references, we can assess how thoroughly the literature review integrates the most recent research and developments in the field.

        To account for both the quality and timeliness of cited references, we propose modifying the Gompertz function to model the reference quality . The Gompertz function is characterized by its sigmoidal, which indicates a slow growth rate at the start and end of a time period, with more rapid growth in the middle phase.  This pattern is often observed in natural phenomena, such as species dynamics~\cite{bruce2016application}, tumor growth~\cite{vaghi2020population}, etc.

        The value of $RQM$ is determined by the average reference quality (ARQ), the shift parameter $\beta$, and the median age of references $S_{mp}$. The calculation procedures of $ARQ$ are as follows: The first step is the extraction of the cited reference list. For most publications, their reference lists could be provided by Semantic Scholar API. For a small number of reviews, the reference list provided by Semantic Scholar may contain errors.  In this case, although there are powerful computer vision-based algorithms~\cite{blecher2023nougat,cheng2023m6doc} available for extracting the reference list within PDFs, our requirements are relatively simple and can be effectively met by relying on the heuristic algorithms or ChatGPT. More specifically, for literature review with a relatively fixed citation format, we can use the PDFMiner~\cite{pdfminer} to read text from PDF files and use heuristic rules to match citations. Alternatively, the text can be analyzed using ChatGPT to extract in-text citations.  The second step is similar to the calculation of the $TNCSI$, where the ChatGPT and a well-designed prompt (as presented in Fig.~\ref{fig:prompt}) are utilized to obtain the topic keyword of the review. Next, we calculate the $TNCSI$ for each reference in the list. To conserve computational resources, we avoid using the ChatGPT to generate keywords for each reference. Instead, the $TNCSI$ of all cited literature is calculated using a sharing topic keyword. Finally, the coverage can be further calculated in Eq.~\eqref{eq:ARQ}:

        \begin{equation}
        \label{eq:ARQ}
            ARQ = \frac{\sum_{i=1}^{N_R} TNCSI(Ref_i)}{N_R},
        \end{equation}
        where $TNCSI(\cdot)$ refers to the $TNCSI$ value of the $i_{th}$ cited reference, and $N_R$ stands for the number of the reference. In certain instances, it has been noted that calculating $TNCSI_s$ for each cited literature is also reasonable. However, this paper primarily emphasizes the current impact of the cited references, hence the utilization of $TNCSI$ in this context.

        The shift parameter $\beta$ can be set empirically or obtained statistically. For statistical estimation, we first examine the distribution of $S_{mp}$ and $ARQ$ of the RiPAMI database. The results indicate that (1) the $S_{mp}$ of over 50\% of the reviews falls within the [5,10] interval, where we denote the lower boundary as $l_s$ and the upper boundary as $r_s$; (2) the average of $ARQ$ is approximately 0.6, denoted as $\overline{ARQ}$. Then, the problem of asserting for $\beta$ is reconceptualized as an optimization problem. As shown in Eq.~\eqref{eq:opt_b}, the objective here is to identify the value of $\beta$ that maximizes the integral of the absolute value of $RQM'(S_{mp}; \beta, \overline{ARQ})$ over the range $l_s$ to $r_s$, subject to the constraints of $\overline{ARQ} = 0.6$. Such an approach would endow $RQM$ with a more discriminative nature. It should be noted that different fields may result in distinct values of $\beta$. For this study, the $\beta$ has been set as 5 for all fields.
        \begin{equation}
            \label{eq:opt_b}
            \beta_{\text{opt}} = \arg\max_\beta \left( \int_{l_s}^{r_s} \left| RQM'(S_{mp}; \beta, \overline{ARQ}) \right| \, dS_{mp} \right).
        \end{equation}

\begin{figure}
             \centering
             \includegraphics[width=0.6\linewidth]{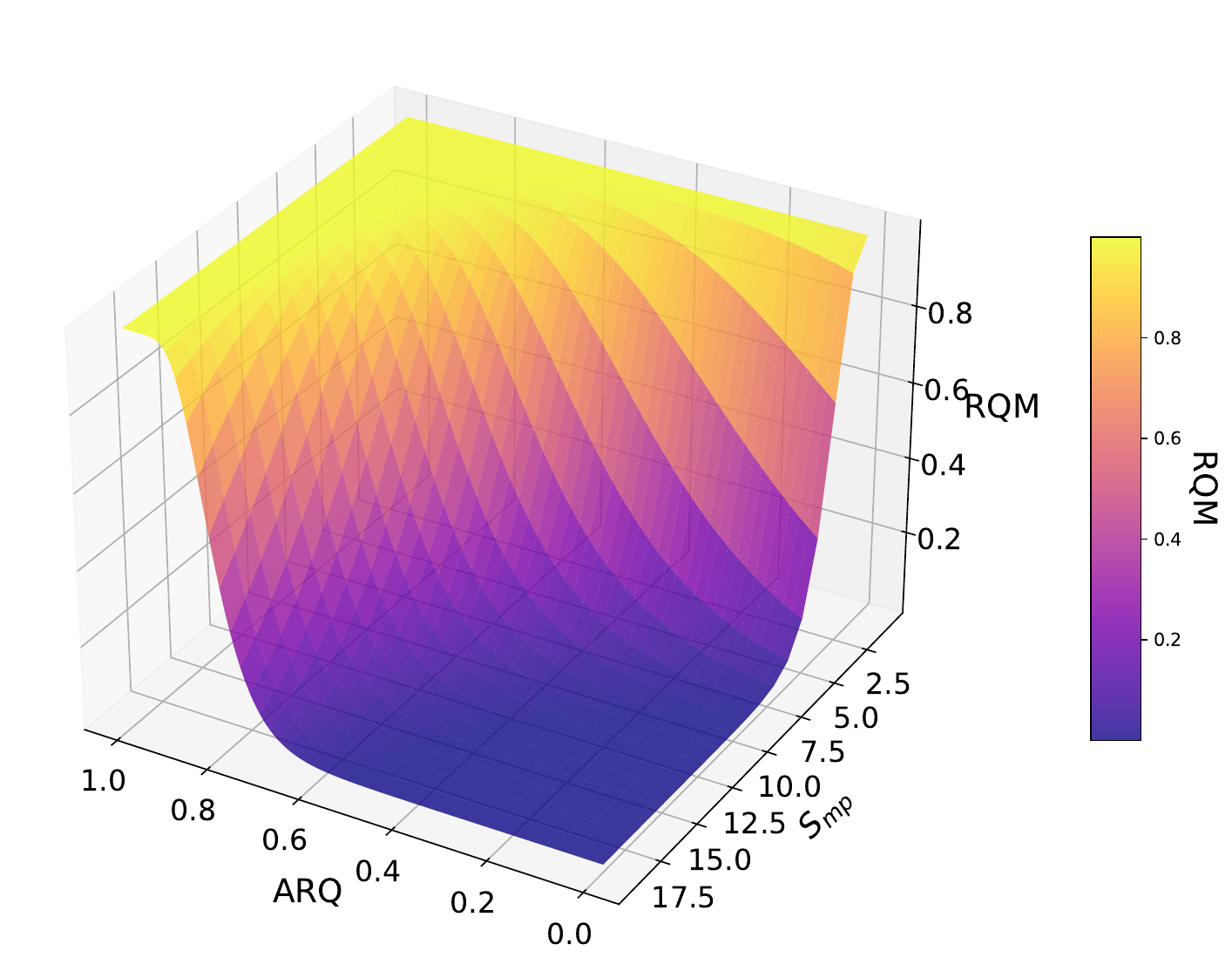}
             \caption{3D Visualization of the Proposed $RQM$}
             \label{fig:vis_RQM}
         \end{figure}

        The range of $RQM$ extends from 0 to 1, where values closer to 1 signify a higher quality of the referenced literature. As illustrated in Fig.~\ref{fig:vis_RQM}, when the $ARQ$ of a paper remains constant, an increase in the variable $S_{mp}$ will lead to a decrease in the $RQM$ value. Conversely, when $S_{mp}$ remains constant, a higher $ARQ$ will elevate the $RQM$ value.

    \subsection{RUI}
    \label{sec:appendix_RUI}
       To evaluate the $RUI$, we may start with the coverage of references before and after publication. This coverage ratio can, to some extent, indicate the extent to which a review requires updating within its field. However, as mentioned earlier, accessing the coverage of a review is difficult. Fortunately, this problem is subtly avoided in calculating the ratio of relevant papers before and after publication. Assuming that the ratio of references containing the topic keyword in the title to all references is $R_k$, the total number of relevant articles can be estimated by dividing the number of articles containing those keywords retrieved from a search engine by $R_k$. Note that $R_k$ generally remains consistent before and after the publication, the Coverage Difference Ratio ($CDR$)  can then be calculated in Eq.~\eqref{eq_CDR}. The theoretical value range of $CDR$ is greater than 0 to positive infinity. When the $CDR$ of a review equals 1, it indicates that the current field has yielded new publications sufficient to constitute half of the literature referenced in the review.
        
       \begin{equation}
        \label{eq_CDR}
            CDR = \frac{N_{pc} \cdot R_k }{R_k \cdot N_{mp}} = \frac{N_{pc}}{N_{mp}},
        \end{equation}
        where $N_{mp}$ and $N_{pc}$ denote the number of relevant literature from the median publication date of the cited references to the publication date of the review, and from the publication date of the review to the current time, respectively.
        
        In addition, similar to the inevitable process of biological aging, literature reviews also undergo a gradual aging process throughout time. Such passage of time bestows upon literature reviews increasing aging progress, where the degree of aging can be conceptualized as a normalized value of the academic impact already achieved. To further explore the aging of reviews in the field of PAMI, we conducted a statistical analysis of the yearly number of newly received citations for reviews published between 2015-2017 in RiPAMI. In contrast to earlier findings, however, the distribution of received citations of reviews over time follows a t-distribution rather than a log-normal distribution of the regular paper, as previously reported in Ref.~\cite{matricciani1991probability,egghe1992citation,moed2005statistical}. Due to the insufficient duration of published sample data, the observation of citation-time trend curves is incomplete. Therefore, we conducted a three-degree polynomial fitting on the limited 6-year citation trend data and transformed the positive segment of the fitted curve into a PDF. To obtain the corresponding cumulative distribution function (CDF), we employed the cumulative trapezoidal numerical integration method for an approximate estimation. 
        Thus, the Review Aging Degree (RAD) is given by:
        \begin{equation}
            RAD(M_{pc}) = \int_{0}^{M_{pc}/12} (px^3 + qx^2 + rx + s) \,dx,
        \end{equation}
        where $M_{pc}$ denotes the duration in months from the publication of the review to the present, $p=-0.003$, $q=0.001$, $r=0.1267$, $s=0.0129$ are the coefficients obtained by polynomial fitting. Please note that the integral symbol used here is for illustrative purposes only. The strict mathematical definition involves the accumulation of discrete trapezoidal areas.

\begin{figure}
             \centering
             \includegraphics[width=0.6\linewidth]{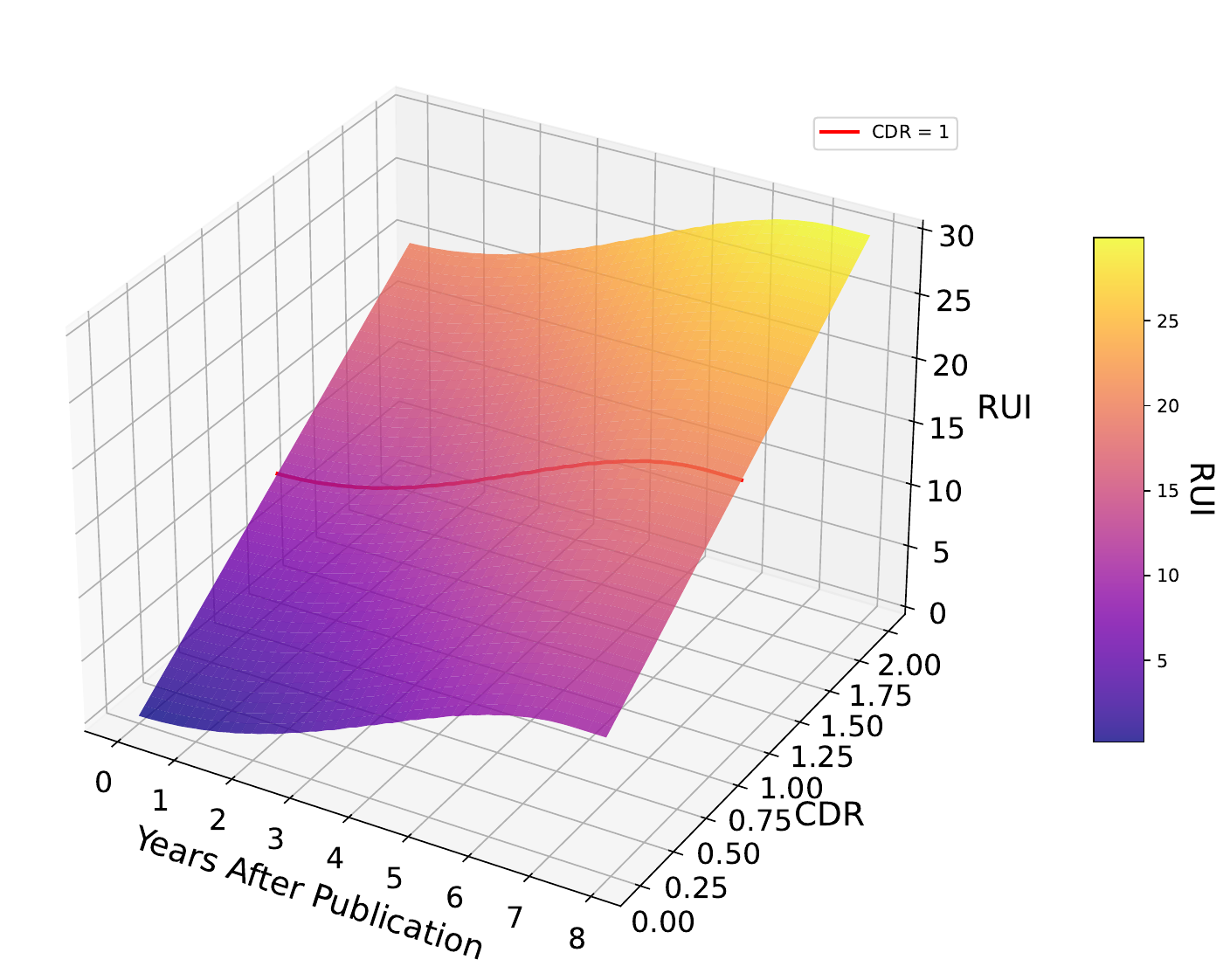}
             \caption{3D Visualization of the Proposed $RUI$}
             \label{fig:vis_RUI}
         \end{figure}

        Finally, the $RUI$ could be obtained by weighted summation of $CDR$ and $RAD$.  A sculptural visualization of the $RUI$'s contours is crystallized in Fig.~\ref{fig:vis_RUI}.

    \subsection{Ensuring Fair Use and Avoiding Metric Misinterpretation}
    \label{sec:metric_fair_use}
    No metric is perfect. Researchers should recognize the limitations of metrics and avoid their misuse or misinterpretation. All citation-based metrics, in particular, are subject to various biases, such as those influenced by the Matthew effect, which can inadvertently contribute to the persistence and dissemination of inaccuracies within academic literature~\cite{Tatsioni2007PersistenceOC}.
    
    \textbf{For TNCSI}. $TNCSI$ is primarily used to assess the cumulative impact of literature reviews and should not be used to evaluate the quality of literature reviews, especially those published in different years.
    
    \textbf{For IEI}. $IEI$ solely represents citation trend changes over a specific past period and should not be extrapolated to predict future citation trends solely.
    
    \textbf{For RQM}. As mentioned earlier, due to practical and conceptual constraints, $RQM$ considers only the average impact and timeliness of references, without accounting for the relevance of the references to the investigated topic. Consequently, there is a potential for manipulating this metric by citing influential references unrelated to the topic to inflate the $RQM$ score.
    
    \textbf{For RUI}. As its name suggests, the $RUI$ is a metric used to assess the extent to which a review requires updating. It is not designed to quantify the degree of obsolescence of the perspectives or conclusions within the review.

    Additionally, it is important to note that all the metrics are designed for article-level analysis of reviews and should not be used to infer the development of an entire field based on the metrics of a single article.
    
    \subsection{Alignment with the Leiden Manifesto}
    
    \begin{table}[ht]
    \renewcommand\arraystretch{1.1}
    \setlength{\tabcolsep}{6pt}
    \begin{center}
        \begin{tabular}{m{0.4\linewidth}c c c c}
            \hline
            Principle & $TNCSI$ & $IEI$ & $RQM$ & $RUI$  \\
            \hline
            \hline
            Quantitative evaluation should support qualitative, expert assessment. & \checkmark & \checkmark & \checkmark & \checkmark \\
            \hline
            Measure performance against the research missions of the institution, group or researcher. & $\times$ & $\times$ & $\times$ & $\times$ \\
            \hline
            Protect excellence in locally relevant research. & N/A & N/A & N/A & N/A \\
            \hline
            Keep data collection and analytical processes open, transparent and simple. & \checkmark & \checkmark & \checkmark & \checkmark \\
            \hline
            Allow those evaluated to verify data and analysis. & \checkmark & \checkmark & \checkmark & \checkmark \\
            \hline
            Account for variation by field in publication and citation practices. & \checkmark & \checkmark & \checkmark & \checkmark \\
            \hline
            Base assessment of individual researchers on a qualitative judgement of their portfolio. & N/A & N/A & N/A & N/A \\
            \hline
            Avoid misplaced concreteness and false precision. & \checkmark & \checkmark & \checkmark & \checkmark \\
            \hline
            Recognize the systemic effects of assessment and indicators. & \multicolumn{4}{c}{\checkmark} \\
            \hline
            Scrutinize indicators regularly and update them. & N/A & N/A & N/A & N/A \\
            \hline
        \end{tabular}
    \end{center}
    \caption{Alignment of Leiden Manifesto Principles with Proposed Metrics: ``\checkmark'' indicates alignment with the principle, ``$\times$'' indicates non-alignment, and ``N/A'' indicates principles not applicable.}
    \label{tab:leiden_alignment}
\end{table}

    Table~\ref{tab:leiden_alignment} shows the alignment of our proposed four metrics with the specific guidelines of the Leiden Manifesto. The metrics—$TNCSI$, $IEI$, $RQM$, and $RUI$—demonstrate strong adherence to most principles outlined in the manifesto. Specifically, all four metrics align with principles emphasizing the integration of quantitative evaluation with qualitative expert assessment, transparency in data collection and analysis, verification of data and analysis by those evaluated, accounting for field-specific variations, and avoiding misplaced concreteness and false precision. 

    The metrics do not align with the principle of measuring performance against the research missions of the institution, group, or researcher, as indicated by the ``$\times$'' in the corresponding row. We would like to point out that this reflects a deliberate trade-off made to balance engineering complexity. While more fine-grained measurements of performance would indeed enhance the objectivity and fairness of the metrics, they would inevitably increase the difficulty of data acquisition. In the future, with advancements in academic search engines and breakthroughs in natural language understanding, this issue is expected to be effectively addressed.

    \subsection{Illustrative Applications of the Indicators}
    \label{sec:app_application_indicators}
    Table.~\ref{tab:eval_main}, Tab.~\ref{tab:eval_main_2} provide more examples of how the four proposed indicators further assist researchers in refining their review selections. Compared to relying solely on titles, citation counts, and publication dates, the proposed metrics provide more information from various perspectives to assist researchers in review selections.

\begin{sidewaystable} 
  \centering 
  \caption{Comparison of Various Reviews with the Proposed Indicators and Metrics (Part~1).}
  \label{tab:eval_main}

  \begin{adjustbox}{width=\textwidth,center}
    \begin{tabular}{|m{0.45\linewidth}llllSSSS|}
    \hline
    \multicolumn{5}{|c|}{Meta-Data} & \multicolumn{4}{c|}{Evaluation} \\
    \cline{1-5} \cline{6-9}
    Title & Year & Cites & Refs  & Topic & {TNCSI↑} & {IEI↑} & {RQM↑} & {RUI↓}\\
    \hline
    \hline
 
        \rowcolor{sblue}
        Object Detection with Deep Learning: A Review~\cite{zhao2019object} & 2018 & 3533 & 252 & deep learning-based objection detection & 1.0 & -7.51 & 0.96 & 99.45\\
        \hline
        \rowcolor{sblue}
        Few-shot Object Detection: a Survey~\cite{antonelli2022few} & 2022 & 34 & 70 & few-shot object detection & 0.26 & -0.58 & 0.68 & 23.49\\
        \hline
        \rowcolor{sblue}
        Recent Few-shot Object Detection Algorithms: A Survey with Performance Comparison~\cite{liu2023recent} & 2022 & 19 & 186 & few-shot object detection & 0.16 & -0.69 & 0.66 & 25.47\\
        \hline
        \rowcolor{sblue}
        Recent progresses on object detection: a brief review~\cite{zhang2019recent} & 2019 & 38 & 110 & object detection & 0.03 & 0.08 & 0.36 & 43.15\\
        \hline
        \rowcolor{sblue}
        A Survey of Modern Deep Learning Based Object Detection Models~\cite{zaidi2022survey} & 2021 & 619 & 113 & object detection & 0.41 & -0.60 & 0.47 & 16.39\\
        \hline
        A Survey on Curriculum Learning~\cite{wang2021survey} & 2021 & 446 & 148 & curriculum learning & 0.92 & 1.09 & 0.14 & 14.25\\
        \hline

        \rowcolor{sblue}
        Automatic Text Summarization Methods: A Comprehensive Review~\cite{sharma2022automatic} & 2022 & 36 & 102 & automatic text summarization & 0.36 & 0.24 & 0.01 & 6.25\\
        \hline
        \rowcolor{sblue}
        Review of Automatic Text Summarization Techniques \& Methods~\cite{widyassari2022review} & 2020 & 181 & 117 & automatic text summarization & 0.90 & 0.32 & 0.06 & 22.10\\
        \hline
        Graph Self-supervised Learning: A Survey~\cite{9770382} & 2021 & 453 & 183 & graph self-supervised learning & 0.99 & -2.85 & 0.95 & 164.52\\
        \hline
        Self-supervised Learning on Graphs: Contrastive, Generative, or Predictive~\cite{wu2021self} & 2021 & 207 & 139 & graph self-supervised learning & 0.90 & -1.02 & 0.94 & 97.67\\
        \hline
        \rowcolor{sblue}
        A Review of Deep-learning-based Medical Image Segmentation Methods~\cite{liu2021review} & 2021 & 436 & 112 & medical image segmentation & 0.75 & -3.11 & 0.26 & 23.36\\
        \hline
        \rowcolor{sblue}
        Biomedical Image Segmentation: A Survey~\cite{alzahrani2021biomedical} & 2021 & 20 & 157 & biomedical image segmentation & 0.13 & -0.28 & 0.00 & 10.11\\
        \hline
        A Survey of Methods, Datasets, and Evaluation Metrics for Visual Question Answering~\cite{sharma2021survey} & 2021 & 33 & 211 & visual question answering & 0.15 & 0.67 & 0.12 & 13.63\\
        \hline
        From Image to Language: a Critical Analysis of Visual Question Answering (VQA) Approaches, Challenges, and Opportunities~\cite{ishmam2023image} & 2023 & 8 & 321 & visual question answering & 0.04 & 0.52 & 0.13 & 3.52\\
        \hline
        \rowcolor{sblue}
        A Survey on Vision Transformer~\cite{9716741} & 2020 & 1537 & 326 & vision transformer & 1.0 & -0.13 & 0.94 & 1246.32\\
        \hline
        \rowcolor{sblue}
        Transformers in Vision: A Survey~\cite{khan2022transformers} & 2021 & 1959 & 285 & transformers in computer vision & 1.0 & -6.41 & 0.98 & 456.40\\
        \hline

    \end{tabular}
  \end{adjustbox}
\end{sidewaystable} 

\begin{sidewaystable} 
  \centering 
  \caption{Comparison of Various Reviews with the Proposed Indicators and Metrics (Part~2, continued).}
  \label{tab:eval_main_2}

  \begin{adjustbox}{width=\textwidth,center}
    \begin{tabular}{|m{0.45\linewidth}llllSSSS|}
    \hline
    \multicolumn{5}{|c|}{Meta-Data} & \multicolumn{4}{c|}{Evaluation} \\
    \cline{1-5} \cline{6-9}
    Title & Year & Cites & Refs  & Topic & {TNCSI↑} & {IEI↑} & {RQM↑} & {RUI↓}\\
    \hline
    \hline
 
        A Survey on Vision Transformer~\cite{9716741} & 2020 & 1537 & 326 & vision transformer & 1.0 & -0.13 & 0.94 & 1246.32\\
        \hline
        \rowcolor{sblue}
        Transformers in Vision: A Survey~\cite{khan2022transformers} & 2021 & 1959 & 285 & transformers in computer vision & 1.0 & -6.41 & 0.98 & 456.40\\
        \hline
        \rowcolor{sblue}
        A Survey of Visual Transformers~\cite{10088164} & 2021 & 248 & 244 & visual transformers & 0.82 & -0.18 & 0.97 & 115.19\\
        \hline
        \rowcolor{sblue}
        A Survey on Efficient Vision Transformers: Algorithms, Techniques, and Performance Benchmarking~\cite{papa2023survey} & 2023 & 18 & 99 & efficient vision transformers & 0.17 & 0.15 & 0.79 & 12.62\\
        \hline
        A Survey on Graph Diffusion Models: Generative AI in Science for Molecule, Protein and Material~\cite{zhang2023survey} & 2023 & 37 & 151 & graph diffusion models & 0.56 & -0.31 & 0.95 & 7.29\\
        \hline
        A Survey on Audio Diffusion Models: Text to Speech Synthesis and Enhancement in Generative AI~\cite{zhang2023survey2} & 2023 & 54 & 141 & audio diffusion models & 0.40 & 0.53 & 0.46 & 49.41\\
        \hline
        Diffusion Models: a Comprehensive Survey of Methods and Applications~\cite{yang2023diffusion} & 2022 & 921 & 394 & diffusion models & 0.91 & -1.26 & 0.72 & 30.98\\
        \hline
        \rowcolor{sblue}
        Geometric Deep Learning on Molecular Representations~\cite{atz2021geometric} & 2021 & 234 & 208 & molecular representations & 0.93 & 1.43 & 0.79 & 23.56\\
        \hline
        Ensemble Deep Learning in Bioinformatics~\cite{cao2020ensemble} & 2020 & 194 & 116 & ensemble bioinformatics & 0.72 & 1.27 & 0.30 & 16.36\\
        \hline
        \rowcolor{sblue}
        A Survey on Self-supervised Learning: Algorithms, Applications, and Future Trends~\cite{gui2023survey} & 2023 & 38 & 308 & self-supervised learning & 0.11 & -0.11 & 0.31 & 11.21\\
        \hline
        \rowcolor{sblue}
        Self-Supervised Learning: Generative or Contrastive~\cite{liu2021self} & 2020 & 1341 & 184 & self-supervised learning & 0.98 & -6.96 & 0.89 & 79.72\\
        \hline
        A Comprehensive Survey on Segment Anything Model for Vision and Beyond~\cite{zhang2023comprehensive}& 2023 & 61 & 223 & segment anything model & 0.95 & 0.06 & 0.97 & 166.69\\
        \hline
        \rowcolor{sblue}
        A Survey on Visual Mamba~\cite{zhang2024survey} & 2024 & 20 & 101 & visual mamba & 0.25 & -2.17 & 0.95 & 45.37\\
        \hline
        A Survey on Large Language Model Based Autonomous Agents~\cite{wang2023survey} & 2023 & 692 & 193 & LLM-based autonomous agents & 1.0 & -0.85 & 0.97 & 261.74\\
        \hline
        Retrieval-Augmented Generation for Large Language Models: A Survey~\cite{gao2023retrieval} & 2023 & 770 & 229 & retrieval-augmented generation & 1.0 & 6.11 & 0.97 & 101.36\\
        \hline

    \end{tabular}
  \end{adjustbox}
\end{sidewaystable} 


\section{Information Extraction Techniques}
\label{sec:appendix_IE_tech}
\subsection{Word Counts}
Word counting differs from character counting, where methods relying solely on regular expressions often yield imprecise results due to the complexity of word segmentation rules. To achieve precise word counting, we utilize the Python NLTK library for word tokenization and exclude all tokens representing non-alphabetic elements. As a result, the presented word count only includes alphabetic words, excluding numbers, contractions, and any special symbols.

\subsection{Visual Elements Counts}
\begin{figure}
     \centering
     \includegraphics[width=0.5\linewidth]{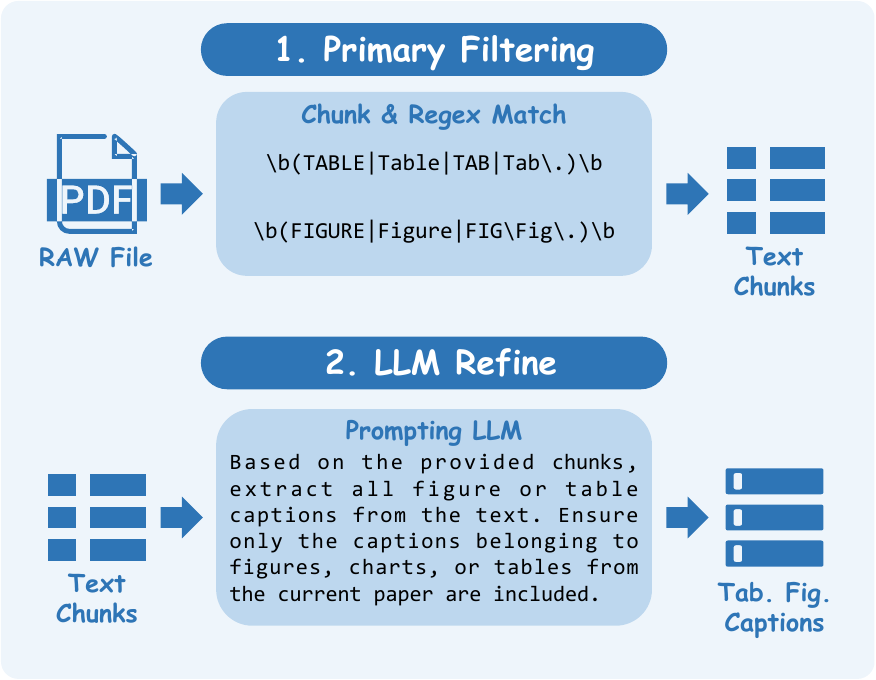}
     \caption{LLM for Figure and Table Caption Extraction: This LLM RAG-based approach allows for the retrieval of figure and table captions at an acceptable cost.}
     \label{fig:fig_tab_caps}
 \end{figure}

The method employed for counting visual elements extends beyond simply relying on the number of layout elements labeled as ``Image'' in the PDF (e.g., ``LTImage'' tag in the PDFMiner). Counting layout elements is prone to inaccuracies, as vector graphics formats like SVG and WMF can encapsulate multiple sub-bitmaps within a single graphic, leading to further errors in the counting process. To address this issue, we utilize LLM-based information extraction techniques to identify the captions associated with all images and tables in the review. The total number of visual elements is determined based on the caption count. As shown in Fig.~\ref{fig:fig_tab_caps}, the first step involves initial filtering of the PDF document to identify chunks containing key terms such as "Fig" and "Tab." The document is first segmented into non-overlapping text chunks, each with a maximum length of 400 characters, using the newline character (`\texttt{\textbackslash n}') as the delimiter. Next, regular expressions are applied within each chunk to filter and retain only those containing the key terms for subsequent analysis. Finally, we utilize the LLM with Retrieval-Augmented Generation (RAG) enabled to analyze these chunks and parse the response into a list of figure and table captions. This method allows for precise extraction of figure and table captions while avoiding the additional costs of full-document processing by the LLM.

\subsection{Review Feature Extraction}
This paper performs a statistical analysis of six features across more than 3,000 review samples, providing insights into factors such as compliance with PRISMA standards and the presence of application sections. Similar to visual element counting, these features are identified using LLM RAG-based information extraction methods.

\begin{figure}
     \centering
     \includegraphics[width=0.5\linewidth]{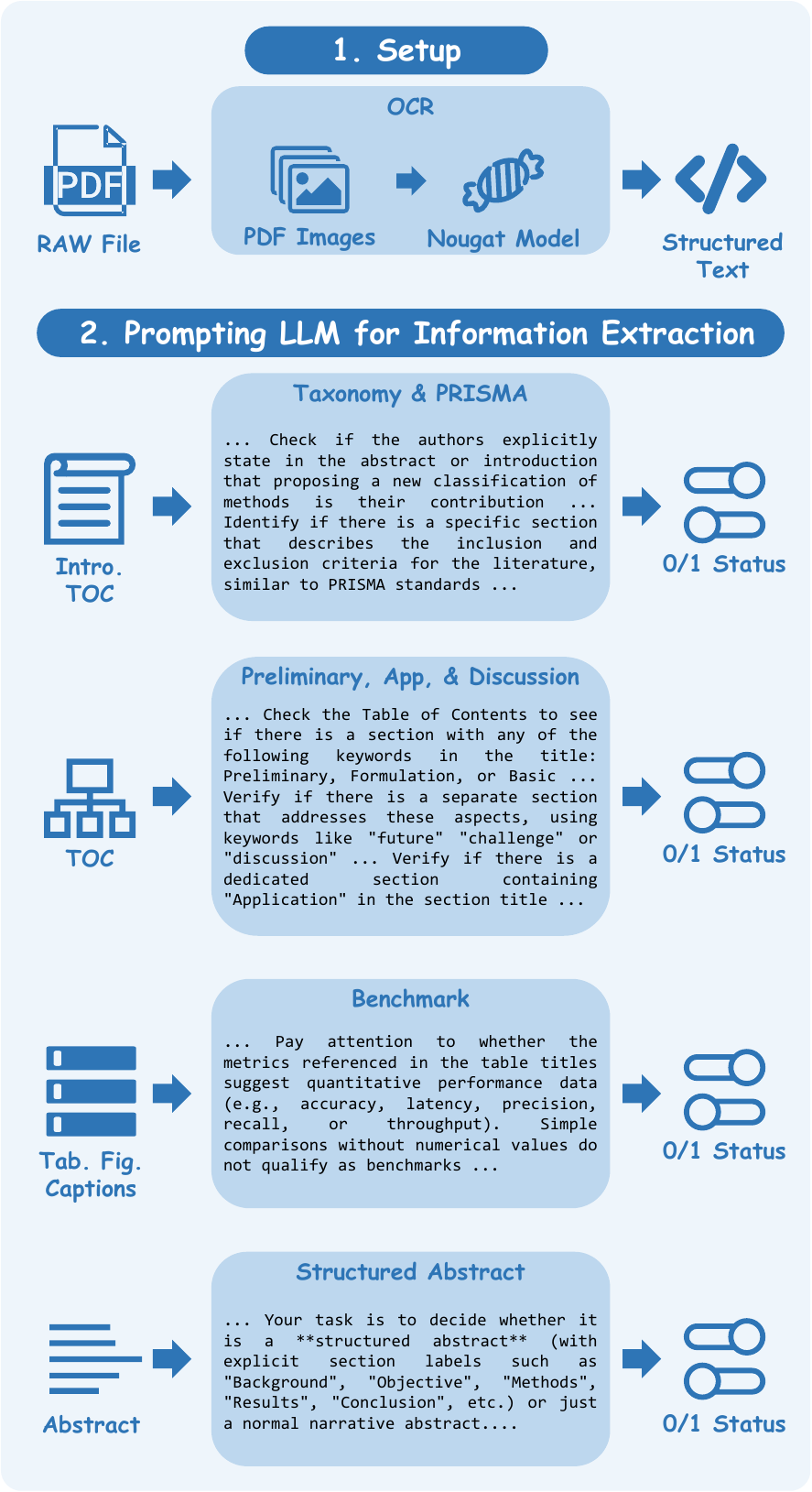}
     \caption{\added{LLM for Review Features Extraction: By utilizing OCR technology~\cite{blecher2023nougat}, we extract structured text and input the introduction section (Intro.), table of contents (TOC), and captions of visual elements into the LLM to derive review features.}}
     \label{fig:six_char_extract}
 \end{figure}

Our proposed method for extracting review features is depicted in Fig.~\ref{fig:six_char_extract}. The process begins with document analysis and recognition. Due to the inherent structure of PDF files, the machine-readable text order often diverges from the natural reading sequence. Additionally, elements such as captions and superscripts can introduce noise, disrupting semantic continuity during direct text extraction. To overcome these challenges, we propose utilizing the Nougat model~\cite{blecher2023nougat} to process document images and convert them into structured text. This structured text enables efficient extraction of the Table of Contents (TOC) and content from individual sections.

We adopt a cost-effective approach during RAG by limiting the need for the LLM to process the entire text. In simple terms, specific sections of the article are selectively provided to the LLM to generate targeted responses. Alongside the title and abstract, the extracted introduction section and TOC are provided to guide the LLM in identifying whether the authors propose a new taxonomy or include a dedicated section for inclusion and exclusion criteria (indicating adherence to PRISMA guidelines) in the reviewed literature. To assess whether the authors discuss preliminaries, applications, or future challenges, we propose relying exclusively on TOC information, further minimizing costs. For benchmarking, the LLM is instructed to analyze extracted captions (illustrated in Fig.~\ref{fig:fig_tab_caps}) to determine whether the authors have performed quantitative benchmarking of existing methods. \added{In addition, the LLM is guided to judge whether the abstract follows structured-abstract standards~\cite{beller2013prisma}.} All responses generated by the LLM are formatted into a structured 0/1 schema and stored in the RiPAMI database.

\subsection{Position Statistics of Section Titles}
\label{sec:appendix_pos_stat_sec_titles}
To analyze the structural characteristics of reviews, we compute the normalized positions of section titles that match a predefined set of keywords (e.g., \emph{introduction}, \emph{conclusion}). The procedure follows three steps.  

First, the keyword list is standardized by removing empty entries, converting to lowercase, and eliminating duplicates while preserving the original order. Second, for each paper, all section titles are retrieved, and the total number of titles is used to normalize their positions within the document. Specifically, the index of each title is divided by $(n-1)$, where $n$ denotes the number of titles in the paper, yielding a value between $0$ (beginning of the document) and $1$ (end of the document). Finally, each title is checked against the keyword list in a case-insensitive substring matching manner, allowing one title to be counted under multiple keywords if applicable.  

This method produces, for each keyword, a distribution of normalized positions across the entire corpus, which forms the basis for our structural analysis.

\section{The Relationship between Review Features and TNCSI}

\begin{figure}
     \centering
     \includegraphics[width=0.5\linewidth]{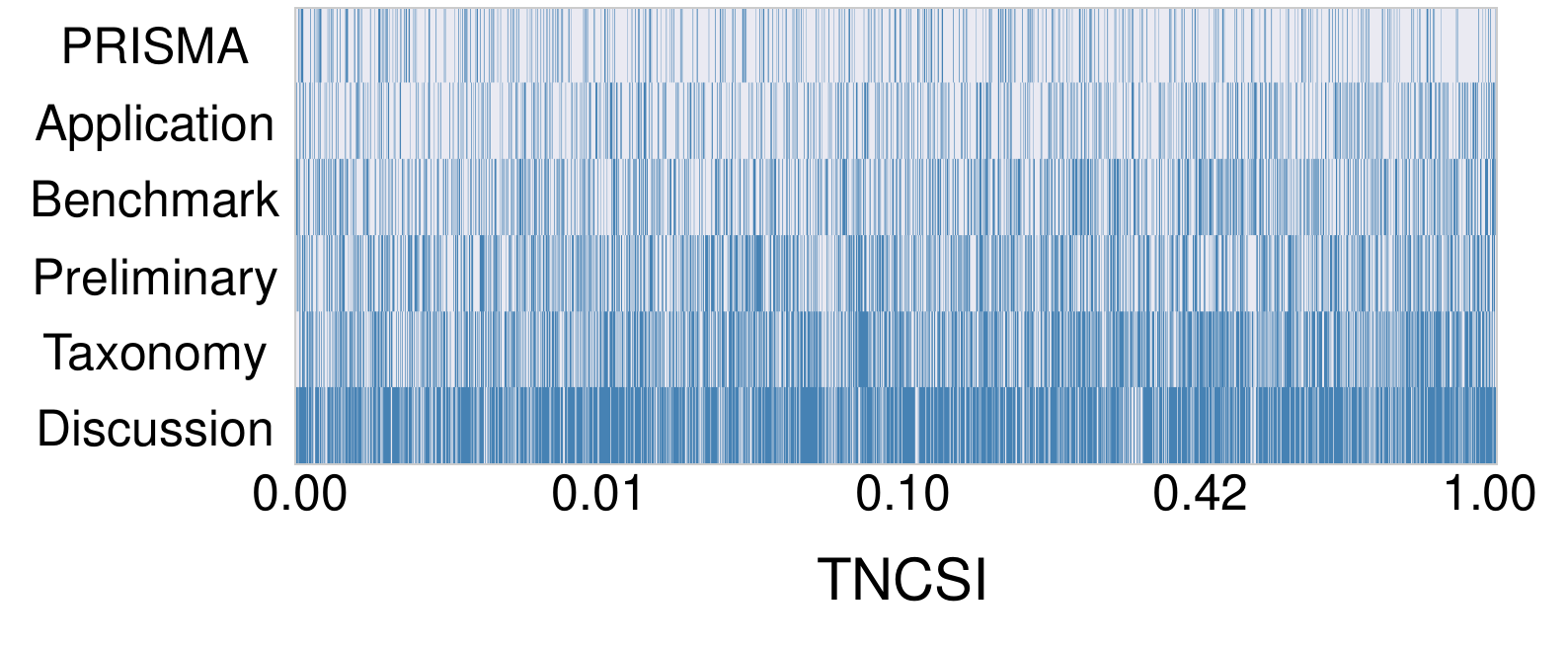}
     \caption{Analysis of Six Review Content Features and Their Association with $TNCSI$: Gene expression heatmap illustrating the independence between six review content features (PRISMA, Application, Benchmark, Preliminary, Taxonomy, and Discussion) and $TNCSI$. Darker shades indicate a match with the feature, corresponding to a value of 1. }
     \label{fig:gene}
\end{figure}

\added{In this study, we dedicate our efforts to analyzing and synthesizing six key types of content features in review articles within the PAMI field. It is important to emphasize that the presence or absence of these features in an article should not be interpreted as a direct measure of its overall quality or potential academic value. As shown in Fig.~\ref{fig:gene}, there is no apparent correlation between the distribution of these six features and $TNCSI$. }

\section{Complete List of Subfields in CV, NLP, and MISC}
\label{sec:appendix_subfields_list}
\added{For transparency and reproducibility, we provide a complete catalogue of subfields across the three major domains, namely CV, NLP, and MISC, as summarized in Table~\ref{tab:subfield_catalogue}.}

\begin{table}[htbp]
    \renewcommand\arraystretch{1.2}
    \setlength{\tabcolsep}{6pt}
    \centering
    \caption{\added{Comprehensive Subfield Catalogue of CV, NLP, and MISC Domains}}
    \label{tab:subfield_catalogue}
    \begin{tabular}{|p{0.12\textwidth}|p{0.82\textwidth}|}
        \hline
        \textbf{Domain} & \textbf{Subfields} \\
        \hline
        CV & action detection, action recognition, activity detection, activity recognition,
        anomaly detection, boundary detection, cnn, computer vision, depth estimation,
        edge detection, emotion recognition, face detection, face recognition,
        facial recognition, gesture analysis, gesture recognition, hand gesture recognition,
        handwriting recognition, human activity recognition, human detection, human pose estimation,
        image captioning, image classification, image clustering, image compression,
        image editing, image enhancement, image generation, image inpainting,
        image matching, image quality assessment, image recognition, image reconstruction,
        image restoration, image retrieval, image segmentation, image-based localization,
        instance segmentation, medical image analysis, medical image segmentation,
        object detection, object tracking, optical character recognition, person re-identification,
        point cloud, saliency detection, salient object detection, scene segmentation,
        scene understanding, super-resolution, superpixels, video object segmentation,
        video processing, video summarization, video understanding, visual question answering,
        visual tracking \\
        \hline
        NLP & dialogue modeling, dialogue systems, document analysis, document analysis and recognition,
        document clustering, document layout analysis, document retrieval, language modeling,
        language modelling, machine translation, named entity disambiguation, named entity recognition,
        natural language processing, question answering, relation extraction, sentiment analysis,
        sentiment classification, text classification, text clustering, text generation,
        text mining, text summarization, text-to-image generation, text-to-speech conversion,
        text-to-speech synthesis \\
        \hline
        MISC & adversarial attack, audio classification, biometric authentication,
        biometric identification, contrastive learning, data mining, data visualization,
        diffusion model, domain adaptation, graph mining, knowledge graph,
        knowledge representation, machine learning interpretability, meta-learning,
        metric learning, multi-label classification, pattern matching, pattern recognition,
        pre-training, pretraining, prompt learning, recommendation systems,
        recommender systems, remote sensing, representation learning,
        self-supervised learning, semantic segmentation, signature verification,
        speech emotion recognition, speech enhancement, speech recognition,
        speech synthesis, speech-to-text conversion, time series analysis,
        time series forecasting, topic detection, topic modeling,
        transfer learning, unsupervised learning, vision language model,
        word embeddings, zero-shot learning \\
        \hline
    \end{tabular}
\end{table}

\section{Code Framework: PyBiblion}
This study involved extensive coding and engineering practices during the processes of dataset construction and statistical analysis. To support this work, we developed an open-source code library, PyBiblion, designed for bibliometric and statistical analysis. As its name suggests, PyBiblion is a Python-based library designed for bibliometric and statistical analysis, drawing from the Greek root ``biblion'', meaning ``book'' or ``document'' to emphasize its focus on scholarly literature.

PyBiblion offers several core advantages, one of which is lazy loading technology. Unlike most existing frameworks, PyBiblion implements lazy loading for paper information. This means that no network communication occurs during the initialization of an instance object; instead, data is only loaded when it is explicitly accessed. This mechanism helps to reduce the number of requests, alleviating server-side communication pressure and fostering a more equitable usage environment. The second advantage of PyBiblion lies in its user-friendliness. Extensive engineering optimizations have been implemented to integrate information retrieval and metric computation seamlessly. This design enables users to compute metrics such as the proposed $TNCSI$ and $RQM$ with a single line of code, offering a highly efficient and intuitive experience. In addition, PyBiblion integrates numerous practical features, including database support, multithreading execution, visualization tools, and statistical analysis capabilities, further enhancing its functionality and versatility.

\end{appendix}
\end{document}